\documentclass{aa}  
\usepackage{dblfloatfix}% http://ctan.org/pkg/dblfloatfix
\usepackage{graphicx}
\usepackage{natbib}
\usepackage{capt-of}
\usepackage{textcomp}
\usepackage{epstopdf}
\usepackage{rotating}
\usepackage{siunitx}
\usepackage{multirow}
\usepackage{hhline}
\usepackage{float}
\floatstyle{plaintop}
\restylefloat{table}
\usepackage{threeparttable}
\usepackage{longtable}
\usepackage{import}
\usepackage{array}
\usepackage{amsmath}
\usepackage{mwe}
\usepackage{subfig}
\usepackage[version=4]{mhchem}
\usepackage[dvipsnames]{xcolor}
\usepackage[colorlinks=true,allcolors=blue]{hyperref}

\begin{document} 

\renewcommand{\thefootnote}{\alph{footnote}}
\renewcommand{\thefootnote}{\fnsymbol{footnote}}

   \title{JWST detections of amorphous and crystalline HDO ice toward massive protostars}
\titlerunning{}

   \author{Katerina Slavicinska\inst{1,2}
            \and
            Ewine F. van Dishoeck\inst{2,3}
            \and
            \L{}ukasz Tychoniec\inst{2}
            \and
            Pooneh Nazari\inst{2,4}
            \and
            Adam E. Rubinstein\inst{5}
            \and
            Robert Gutermuth\inst{6}
            \and
            Himanshu Tyagi\inst{7}
            \and
            Yuan Chen\inst{2}
            \and
            Nashanty G. C. Brunken\inst{2}
            \and
            Will R. M. Rocha\inst{2}
            \and
            P. Manoj\inst{7}
            \and
            Mayank Narang\inst{8}
            \and
            S. Thomas Megeath\inst{9}
            \and
            Yao-Lun Yang\inst{10}
            \and
            Leslie W. Looney\inst{11, 12}
            \and
            John J. Tobin\inst{12}
            \and
            Henrik Beuther\inst{13}
            \and
            Tyler L. Bourke\inst{14}
            \and
            Harold Linnartz\inst{1}
            \and
            Samuel Federman\inst{9}
            \and
            Dan M. Watson\inst{5}
            \and
            Hendrik Linz\inst{13,17}
            }

    \institute{\textit{affiliations can be found after the references}}

   \date{Received 28 February 2024 / Accepted 23 April 2024}

% \abstract{}{}{}{}{} 
% 5 {} token are mandatory

% \abstract{}{}{}{}{} 
% 5 {} token are mandatory

  \abstract
  % context heading (optional)
   {Tracing the origin and evolution of interstellar water is key to understanding many of the physical and chemical processes involved in star and planet formation. Deuterium fractionation offers a window into the physicochemical history of water due to its sensitivity to local conditions.}
  % aims heading (mandatory)
   {This work aims to utilize the increased sensitivity and resolution of the \textit{James Webb} Space Telescope (JWST) to quantify the HDO/H$_{2}$O ratio in ices toward young stellar objects (YSOs) and to determine if the HDO/H$_{2}$O ratios measured in the gas phase toward massive YSOs (MYSOs) are representative of the ratios in their ice envelopes.}
  % methods heading (mandatory)
   {Two protostars observed in the Investigating Protostellar Accretion (IPA) program using JWST NIRSpec were analyzed: HOPS 370, an intermediate-mass YSO (IMYSO), and IRAS 20126+4104, a massive YSO (MYSO). The HDO ice toward these sources was quantified via its 4.1 $\mu$m band. The contributions from the CH$_{3}$OH combination modes to the observed optical depth in this spectral region were constrained via the CH$_{3}$OH 3.53 $\mu$m band to ensure that the integrated optical depth of the HDO feature was not overestimated. H$_{2}$O ice was quantified via its 3 $\mu$m band. New laboratory IR spectra of ice mixtures containing HDO, H$_{2}$O, CH$_{3}$OH, and CO were collected to aid in the fitting and chemical interpretation of the observed spectra.}
  % results heading (mandatory)
   {HDO ice is detected above the 3$\sigma$ level in both sources. It requires a minimum of two components, one amorphous and one crystalline, to obtain satisfactory fits. The H$_{2}$O ice band at 3 $\mu$m similarly requires both amorphous and crystalline components. The observed peak positions of the crystalline HDO component are consistent with those of annealed laboratory ices, which could be evidence for heating and subsequent re-cooling of the ice envelope (i.e., thermal cycling). The CH$_{3}$OH 3.53 $\mu$m band is fit best with two cold components, one consisting of pure CH$_{3}$OH and the other of CH$_{3}$OH in an H$_{2}$O-rich mixture. From these fits, ice HDO/H$_{2}$O abundance ratios of 4.6$\pm$1.8$\times$10$^{-3}$ and 2.6$\pm$1.2$\times$10$^{-3}$ are obtained for HOPS 370 and IRAS 20126+4104, respectively.}
  % conclusions heading (optional)
   {The simultaneous detections of both crystalline HDO and crystalline H$_{2}$O corroborate the assignment of the observed feature at 4.1 $\mu$m to HDO ice. The ice HDO/H$_{2}$O ratios are similar to the highest reported gas HDO/H$_{2}$O ratios measured toward MYSOs as well as the hot inner regions of isolated low-mass protostars, suggesting that at least some of the gas HDO/H$_{2}$O ratios measured toward massive hot cores are representative of the HDO/H$_{2}$O ratios in ices. The need for an H$_{2}$O-rich CH$_{3}$OH component in the CH$_{3}$OH ice analysis supports recent experimental and observational results that indicate that some CH$_{3}$OH ice may form prior to the CO freeze-out stage in H$_{2}$O-rich ice layers.}


   \maketitle
%
%________________________________________________________________

\section{Introduction}

Water is one of the most abundantly detected interstellar molecules in both the gas and solid phase, and it plays countless essential roles in physical and chemical processes involved in the birth and evolution of stars, planets, and life as we know it (e.g., \citealt{van2013interstellar,van2021water}). Despite this significance and ubiquity, its formation and evolution in the star formation cycle are still not fully understood.

Detections of abundant water ice toward molecular clouds and dense prestellar cores show that significant quantities of water can be produced in the solid state even before the earliest stages of star formation \citep{tanaka1990three,knez2005spitzer,boogert2011ice,boogert2013infrared,mcclure2023ice}. In these environments, water forms on the surfaces of silicate and carbonaceous interstellar grains via cold atom addition chemistry along with other simple species like CO$_{2}$, NH$_{3}$, and CH$_{4}$, resulting in the growth of ice layers that can be observed via infrared spectroscopy (\citealt{boogert2015observations} and references therein). There is mounting evidence that these water-rich prestellar ice mantles also serve as the birthplace and reservoir of many other more complex molecules, some of which could have prebiotic relevance \citep{herbst2009complex,jorgensen2020astrochemistry,nazari2022n,mcclure2023ice,chen2023coccoa,rocha2023jwst,nazari2024hunt}.

However, it is uncertain how much of this primordial water ice and its associated chemical complexity survive the subsequent intense thermal and radiative processes that occur upon the formation of protostars and protoplanetary disks. Once a protostellar source forms from a prestellar core and begins to generate radiation, it can cause its surrounding water ice to sublimate into the gas phase via photo- or thermal desorption, where it may be destroyed and subsequently reformed via hot gas-phase neutral-neutral reactions \citep{bergin1998postshock,harada2010new,visser2011chemical,furuya2013water,van2014water,owen2015astro}. The extent to which such processing of the prestellar water ice occurs within the hot, radiation-dominated regions of protostellar envelopes and protoplanetary disks has important implications for not only the physicochemical processes that drive their evolution, but also for the chemical inventory available to the planetesimals that eventually begin to form and grow in these environments (i.e., the question of \textit{inheritance versus reset}, \citealt{van2017astrochemistry,oberg2021astrochemistry}).

Quantifying the deuterium fractions (D/H ratios) of water offers a promising path to understanding the extent of this processing because deuterium fractionation reactions are extremely sensitive to the physicochemical conditions in which they occur. Therefore, a molecule's D/H ratios reflect the environment in which it formed and existed. In molecular clouds and prestellar cores, molecules are thought to be deuterium enriched due to an increase in the relative abundance of available atomic D (with respect to atomic H) caused by the enhanced gas-phase formation of H$_{2}$D$^+$ \citep{roberts2003enhanced}:\\

\noindent \ce{H3+ + HD $\rightleftharpoons$ H2D+ + H2 + 232 K}.\\

At the low temperatures relevant to ice formation ($T<$50 K), this reaction mostly proceeds in the forward direction due to its energy barrier. The following dissociative recombination reaction can then free D atoms, increasing deuterium atoms that can land on cold dust grains and react to form a higher abundance of deuterated molecules \citep{tielens1983surface}:\\

\noindent \ce{H2D+ + e- $\rightarrow$ H2 + D}\\

Water formed in cold prestellar environments is therefore expected to be deuterium enriched relative to bulk interstellar and solar D/H values (2.0$\times$10$^{-5}$, corresponding to an HDO/H$_{2}$O ratio of 4$\times$10$^{-5}$ if there is no fractionation, \citealt{prodanovic2010deuterium,geiss2003isotopic}). The relative abundance of H$_{2}$D$^+$ is further enhanced in the absence of neutral species like O and CO that destroy H$_{3}$$^+$ and its isotopologs, so deuterium enrichment is expected to be even greater in the molecules formed during the coldest ($<$20 K) and densest stages of prestellar cores, when O and CO are depleted from the gas phase as they freeze out on the grains.

Interstellar HDO was first detected in the gas phase in the massive star-forming region Orion KL \citep{turner1975microwave}. The relatively high brightness of the hot ($>$100 K) inner cores of such massive young stellar objects (MYSOs) allows multiple lines of HDO to be detected relatively easily with millimeter spectroscopy. The detection of these lines, along with the lines of H$_{2}^{18}$O, a water isotopolog that is usually optically thin, has enabled the determination of the gas HDO/H$_{2}$O ratios toward many of these objects in the last few decades. These ratios typically range between 10$^{-4}$ to a few 10$^{-3}$ \citep{helmich1996excitation,van2006water,neill2013abundance,emprechtinger2013abundance,coutens2014water}, an enhancement of up to two orders of magnitude relative to the cosmic standard D/H. Highly sensitive millimeter interferometers also enable reliable measurement of gas HDO/H$_{2}$O ratios in the hot inner regions of the much dimmer low-mass young stellar objects (LYSOs), whose ratios see a similar enhancement \citep{persson2013warm,persson2014deuterium,jensen2019alma,jensen2021alma}.

This enhancement has been interpreted as evidence that at least some of the water gas detected in protostars has its origins in prestellar ices, in line with the assumption that many of the gas-phase species detected in protostellar hot cores and corinos come directly from ices via thermal desorption. Additionally, the gas HDO/H$_{2}$O ratios of clustered LYSOs are remarkably similar to those measured in comets, suggesting that prestellar water ice and its associated chemical complexity may be largely inherited by the planetesimals that grow in the outer protoplanetary disk \citep{persson2014deuterium}.

While studying HDO/H$_{2}$O ratios in the gas phase has clearly improved our understanding of the chemical history of water throughout star formation across a range of stellar masses, a glaring question remains: are the HDO/H$_{2}$O ratios measured in protostellar hot gas actually representative of the protostellar HDO/H$_{2}$O ice ratios, or has the detected water gas already experienced substantial alteration by gas-phase reactions? To answer this question, the HDO/H$_{2}$O ratio in ices must be probed directly.

Despite H$_{2}$O being the most abundant observable hydrogen-bearing ice by a large margin, a secure detection of its deuterated isotopologs is an exceptional observational challenge. Because it begins to form in the earliest stages of cloud formation, its D/H ratio in the gas phase is a couple orders of magnitude below that of molecules like CH$_{3}$OH and H$_{2}$CO thought to form only in the later colder stages of cloud formation \citep{cazaux2011interstellar,taquet2012formaldehyde,caselli2012our,ceccarelli2014deuterium,furuya2016reconstructing}, so the column density of HDO ice is expected to be very low. Furthermore, the strongest IR band of HDO, its O-D stretching mode at $\sim$4.1 $\mu$m ($\sim$2440 cm$^{-1}$), lies next to the short-wavelength wing of the strong asymmetric C=O stretching mode of CO$_{2}$ ice, which sometimes warps the continuum due to grain shape effects and complicates the detection and quantification of spectrally adjacent species \citep{dartois2022influence,dartois2024spectroscopic}. In the amorphous state, the O-D stretch of HDO is also broad and relatively weak \citep{dartois2003revisiting,galvez2011hdo}.

The first tentative HDO ice detection at 4.1 $\mu$m was reported in data from the Infrared Space Observatory (ISO) more than two decades ago by \cite{teixeira1999discovery}, but the detection was soon called into question by \cite{dartois2003revisiting}, who pointed out, among several other concerns, that the observed absorption could also be attributed to a weak CH$_{3}$OH combination mode whose wavelength overlaps with that of the HDO O-D stretch. They went on to report HDO upper limits ranging from 0.16-1.1\% with respect to H$_{2}$O toward four intermediate-mass and massive young stellar objects (IMYSOs and MYSOs) with low CH$_{3}$OH column densities. \cite{parise2003search} expanded the search for HDO ice to low-mass young stellar objects (LYSOs) and also derived upper limits, ranging from 0.5-2\% with respect to H$_{2}$O. Almost a decade later, \cite{aikawa2012akari} reported tentative detections and measured very high HDO/H$_{2}$O ratios (2-22\%) toward four LYSOs, but they cautioned that the HDO ice models did not provide a perfect fit to the observations, and their analysis did not take into account contribution from CH$_{3}$OH ice to the HDO ice spectral region. Furthermore, given the recent characterization of how grain growth affects the continuum around the CO$_{2}$ ice feature \citep{dartois2022influence}, it is possible that some of the observed "broad absorption" that was attributed to HDO ice by \cite{aikawa2012akari} was caused by a warping of the local continuum via grain shape effects, particularly in the case of IRAS 04302+2247.

It is evident that the infrared telescopes capable of observing ices in the 4 $\mu$m range thus far have lacked sufficient sensitivity to conclusively detect solid-state HDO. The unprecedented sensitivity of the Near Infrared Spectrograph (NIRSpec) instrument on the recently launched \textit{James Webb} Space Telescope (JWST) provides a prime opportunity to revisit the search for interstellar HDO ice.

Here we report HDO ice detections toward two protostars, HOPS 370, and IRAS 20126+4104 (hereafter IRAS 20126), observed with JWST's NIRSpec integral field unit (IFUs) as part of the GO program 1802, Investigating Protostellar Accretion (IPA) across the mass spectrum (Program ID 1802, PI Tom Megeath; \citealt{megeath2021investigating}). Care is taken to ensure that the observed HDO ice detections cannot be attributed to the CH$_{3}$OH combination mode at 4 $\mu$m by establishing a CH$_{3}$OH column density via fitting high-resolution laboratory spectra to the 3.53 $\mu$m band. Additional high-resolution laboratory spectra of HDO are collected to enable the fitting of the observational spectra and subsequently derive HDO column densities. The laboratory spectra measured and used in this work are made publicly available via the Leiden Ice Database (LIDA, \citealt{rocha2022lida}). Comparisons are made between the derived HDO ice abundances to those measured in the gas phase in the interstellar medium. The resulting implications for the chemical evolution of water throughout the star formation process are then discussed.

\section{Methods}

\subsection{Investigated sources}
\label{txt:investigated_sources}

\begin{table*}[h!]
\caption{Source parameters and aperture coordinates used to extract spectra from the NIRSpec IFU datacubes.}
\begin{center}
\begin{tabular}{c c c c c c}
\hline
        Source & Size & Distance (pc) & $L$ ($L_{\odot}$) & RA & Dec\\
        \hline
        HOPS 370 & IMYSO & 390$^{(a)}$ & 314$^{(b)}$ & 05:35:27.6417 & -05:09:33.944\\
        IRAS 20126+4104 & MYSO & 1550$^{(c)}$ & 1.1$\times$10$^{4}$$^{(d)}$ & 20:14:26.2358 & +41:13:32.445\\
    \hline
     %   \noalign{\smallskip}
     \label{tab:sources}
\end{tabular}

\begin{tablenotes}
$^{(a)}$ \citet{tobin2020vla2}. $^{(b)}$ \citet{tobin2020vla}. $^{(c)}$ \citet{reid2019trigonometric}. $^{(d)}$ \citet{cesaroni2023herschel}
\end{tablenotes}

\end{center}
\end{table*}

\begin{figure}[h]
\centering
\includegraphics[width=\linewidth]{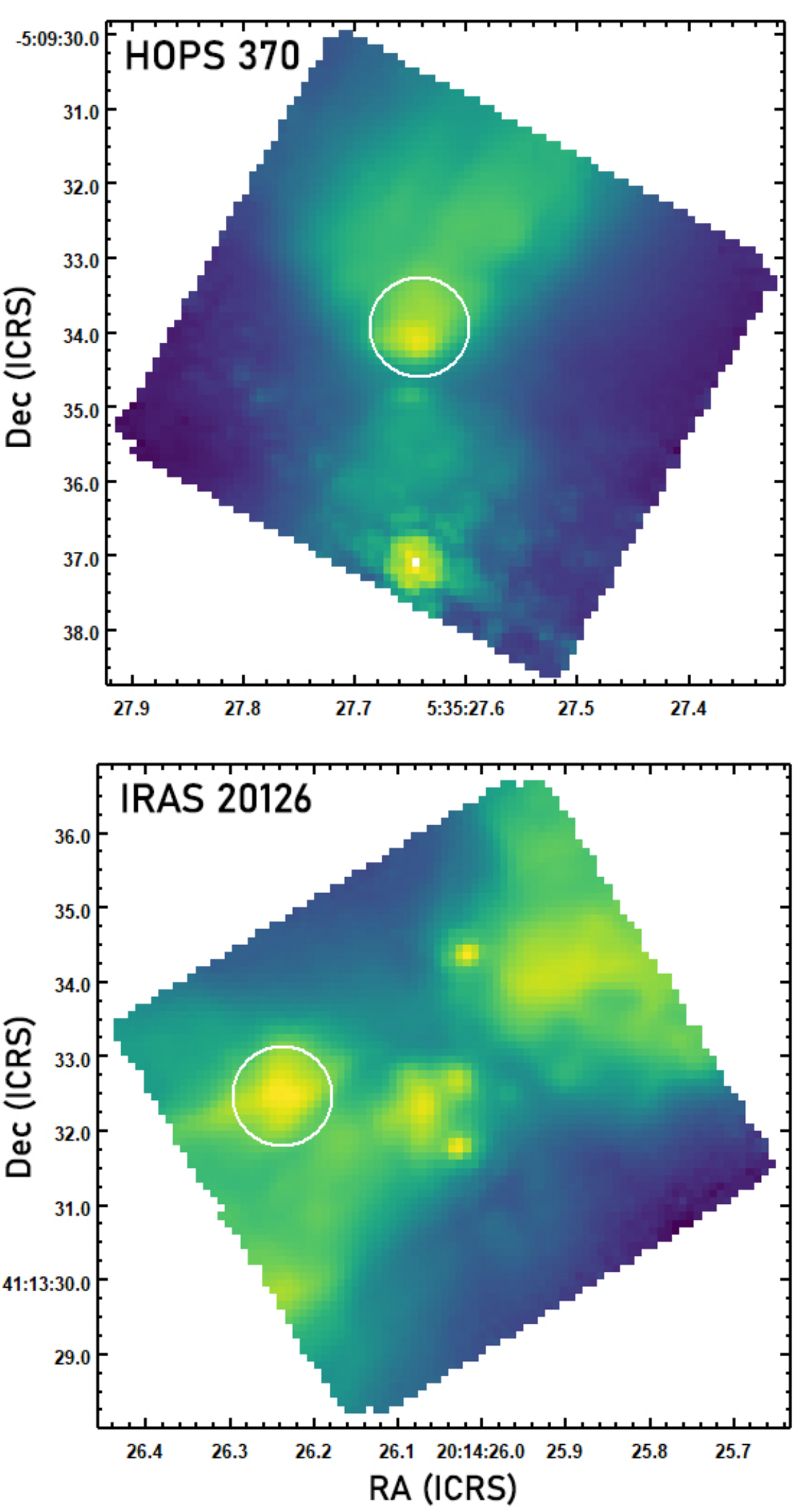}
\caption{5 $\mu$m continuum maps with the aperture used to extract the analyzed spectra at 5 $\mu$m shown in white (7$\times$ diffraction-limited radius at 5 $\mu$m=0.678").}
\label{fig:datacubes}
\end{figure}

The IPA program observed five YSOs in their primary accretion phase, with their bolometric luminosities spanning approximately six orders of magnitude, using the NIRSpec IFU \citep{federman2023investigating}. Out of these sources, two are presently analyzed due to their high S/N and because their IFUs contain lines of sight that have a non-saturated 3 $\mu$m O-H stretching feature of H$_{2}$O ice, enabling quantification of its column density and characterization of its morphology. Table~\ref{tab:sources} presents an overview of these sources.

HOPS 370 is an IMYSO with a spectral energy distribution (SED) near the Class 0/Class I border located within the very clustered OMC-2 star-forming region in the Orion A molecular cloud. It is likely currently experiencing a high accretion rate, and its compact complex organic molecule (COM) emission is similar to those of hot corinos \citep{tobin2020vla}. It is driving a powerful bipolar outflow observed in the far IR, millimeter, and centimeter wavelength ranges \citep{shimajiri2008millimeter,gonzalez2016herschel,osorio2017star,sato2023alma}.

IRAS 20126 is a deeply embedded MYSO, the most massive and luminous protostar of the sample. The source contains a hot and massive disk \citep{chen2016hot} as well as jet-driven outflows \citep{caratti2008molecular}. Although initial observations suggested that its environment is unusually isolated for a MYSO \citep{qiu2008spitzer}, later observations revealed the presence of a young cluster within a 1.2 pc vicinity \citep{montes2015x}. Two likely companion YSOs are present within the 6"x6" FOV in the NIRSpec IFU of this source \citep{federman2023investigating}.

\subsection{Observations and data reduction}

The investigated spectra were collected with the NIRSpec IFU using the G395M grating ($R$=$\lambda / \Delta \lambda$$\sim$700-1300, \citealt{rubinstein2023ipa}), providing data from 2.9-5.1 $\mu$m (3450-1960 cm$^{-1}$) with no detector gaps, unlike the higher-resolution G395H grating, which has a detector gap in the IFU at 4.1 $\mu$m, where the strongest HDO IR band is found. A 4-point dither pattern was used to obtain a 6"$\times$6" mosaic with a 0.2" spatial resolution. Detailed descriptions of the data reduction are provided in \citet{federman2023investigating} and \citet{narang2024discovery}.

The regions in the IFUs from which spectra were extracted for analysis were chosen by searching through the datacube for the highest S/N lines of sight in which the H$_{2}$O 3 $\mu$m feature is not saturated or extincted. In both cases, the ice features in the selected positions are from the cold envelopes seen in projection in front of bright scattered light emission from the protostars. The HOPS 370 spectrum was extracted near the source center over a knot of shocked gas. Spectra extracted near the central position of IRAS 20126 cannot be used for the analysis in this work because the flux at 3 $\mu$m at these coordinates is so low that the profile of the H$_{2}$O feature is contaminated by an extended weak emission that may originate from a photodissociation region along the line of sight. Extended weak emission is present toward all of the IPA sources and is seen most distinctly at the edges of the datacubes, where the source continuum fluxes are minimal and a weak PAH emission feature at 3.3 $\mu$m is clearly observed. The analyzed spectrum of IRAS 20126 is therefore extracted from a bright position toward an outflow cavity, where the PAH emission flux is negligible relative to the overall observed flux.

The coordinates of the apertures used to extract the spectra are provided in Table~\ref{tab:sources} and are shown on the IFU continuum maps in Figure~\ref{fig:datacubes}. A wavelength-dependent extraction aperture of 7 times the diffraction-limited radius was used, which results in an aperture that is $>$3 times the full width at half maximum (FWHM) of the NIRSpec IFU point spread function over the entire spectrum. Because the investigated sources have small radial velocities relative to the spectral resolution, the final extracted spectra are not corrected for source velocity \citep{brunken2024jwst}.

\subsection{Gas emission subtraction}
Both spectra contain gas emission features, including those of H$_{2}$, H I, [Fe II], and CO \citep{federman2023investigating,narang2024discovery,rubinstein2023ipa}. The strongest of these features were subtracted out of the spectral regions of interest (3.33-4.18 $\mu$m) via Gaussian fits to prevent their interference with the fitting functions used to analyze the ice features. Most of the gas emission lines are well-resolved from other neighboring gas lines and thus required only single Gaussian profiles to be subtracted out. The few emission lines that are blended were subtracted out using multiple Gaussian profiles.

In all subsequent spectral figures, the original spectra before gas emission subtraction are plotted in gray, and the emission-subtracted spectra are over-plotted in color.

\subsection{Laboratory data collection}
\label{txt:lab_data}
High-resolution (0.5 cm$^{-1}$) spectra of various laboratory ice mixtures containing HDO, H$_{2}$O, CO, and CH$_{3}$OH ranging from temperatures of 15-150 K were collected on the InfraRed Absorption Setup for Ice Spectroscopy (IRASIS) in the Leiden Laboratory for Astrophysics for use in fitting the observed spectra. A schematic of the set-up can be found in \cite{rachid2021infrared}, and recent upgrades to the set-up are described in \cite{slavicinska2023hunt}. The method used to create ice mixtures with specific ratios is similar to that described in \cite{yarnall2022new}. The chemical compositions and mixing ratios of the laboratory ice mixtures are presented in Table~\ref{tab:lab_spectra}. Additional experimental details can be found in Appendix~\ref{app:lab}.

\begin{figure*}[h!]
\centering
\includegraphics[width=\linewidth]{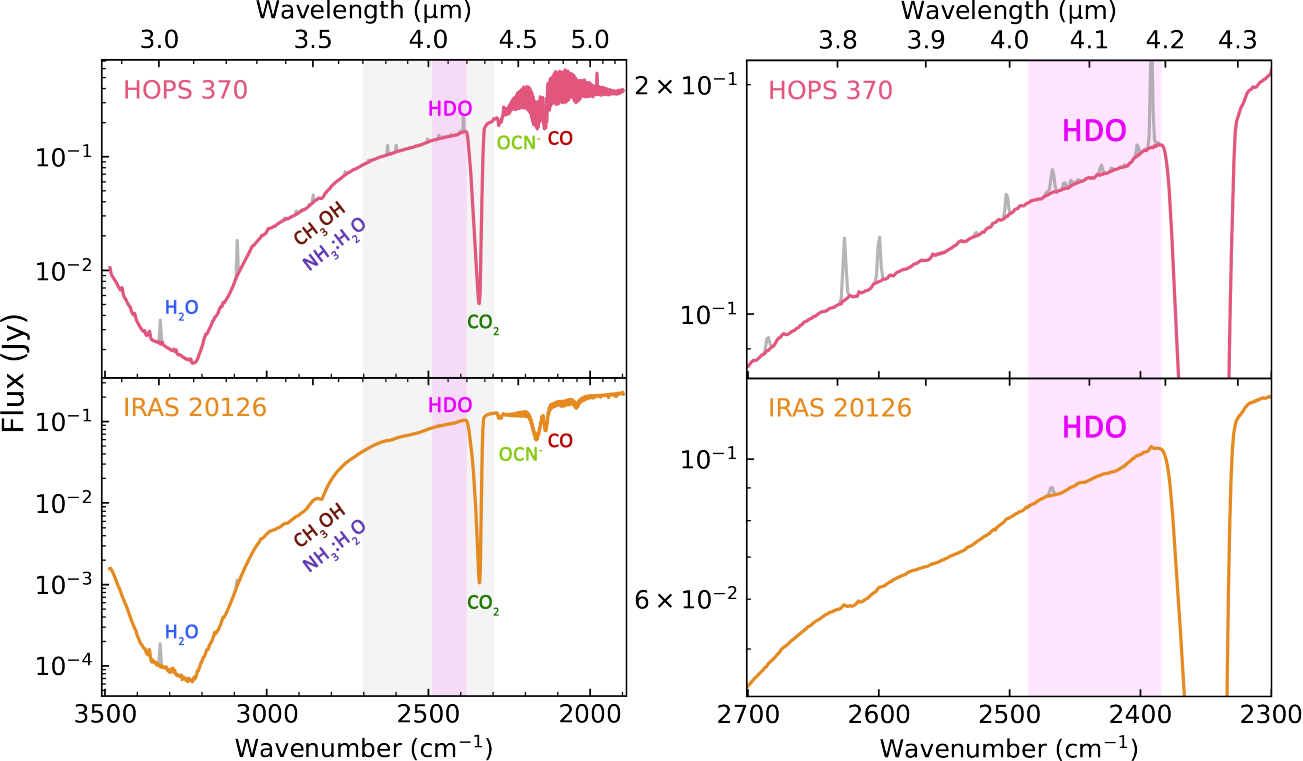}
\caption{Extracted spectra of the investigated sources. Left: full NIRSpec spectra with major ice features labeled. The original spectra before gas emission subtraction are plotted in gray. The spectral region of the strongest HDO absorption, the O-D stretching mode, is indicated via magenta shading. The spectral region plotted on the right is indicated via gray shading. Right: a zoomed-in view of the spectral region of interest for HDO ice.}
\label{fig:all_spectra}
\end{figure*}

\begin{table}
\caption{Laboratory spectra collected for this work.}
\begin{center}
\begin{tabular}{c c}
\hline
        Mixture & Ratio \\
        \hline
        \multirow{2}{*}{HDO:H$_{2}$O} & 0.4:100 (thick)$^{*}$\\
        & 0.4:100 (thin)\\
        \hline
        CH$_{3}$OH & pure \\
        \hline
        \multirow{4}{*}{CH$_{3}$OH:H$_{2}$O} & 2:1 \\
        & 1:1 \\
        & 1:2 \\
        & 1:5 \\
        \hline
        \multirow{4}{*}{CH$_{3}$OH:CO} & 2:1 \\
        & 1:1 \\
        & 1:2$^{*}$ \\
        & 1:5$^{*}$ \\
    \hline
     %   \noalign{\smallskip}
     \label{tab:lab_spectra}
\end{tabular}

\begin{tablenotes}
    \item[\emph{}]{$^*$ indicates that the strongest peak of the primary matrix component is saturated.}
\end{tablenotes}

\end{center}
\end{table}

All of the spectra are made available for public use and enjoyment via LIDA \citep{rocha2022lida}. As much of this work focuses on fitting very weak features (i.e., the O-D stretch of highly diluted HDO at 4.1 $\mu$m and the combination modes of CH$_{3}$OH at 3.9 $\mu$m), the strongest peak of the primary ice matrix component (i.e., the C$\equiv$O stretching mode of CO or the O-H stretching mode of H$_{2}$O) was allowed to saturate in three of the experiments so that the weaker ice features have sufficient S/N for accurate fitting. The spectra with such a saturated peak are indicated with an asterisk (*) in Table~\ref{tab:lab_spectra}. We strongly warn against using the saturated peaks in these spectra to fit observational data.

The peak positions, FWHMs, and apparent band strengths of the O-D stretching mode from 15 to 150 K were extracted from the HDO:H$_{2}$O mixtures for both the amorphous and crystalline phases and are presented in Table~\ref{tab:hdo_peak_char} and Figures~\ref{fig:hdo_band_strength} and \ref{fig:hdo_pos_fwhm}. As discussed in detail in Section~\ref{txt:hdo_ice}, the variation in amorphous HDO band strength with respect to temperature has particularly important ramifications on the derived amorphous HDO abundances. The peak positions of the CH$_{3}$OH 3.53 $\mu$m band from 15 to 150 K in all of the CH$_{3}$OH-containing spectra are presented in Tables~\ref{tab:ch3oh_peak_char_h2o} and~\ref{tab:ch3oh_peak_char_co} and Figure~\ref{fig:ch3oh_pos}.

\section{Results}

\begin{figure*}[h!]
\centering
\includegraphics[width=\linewidth]{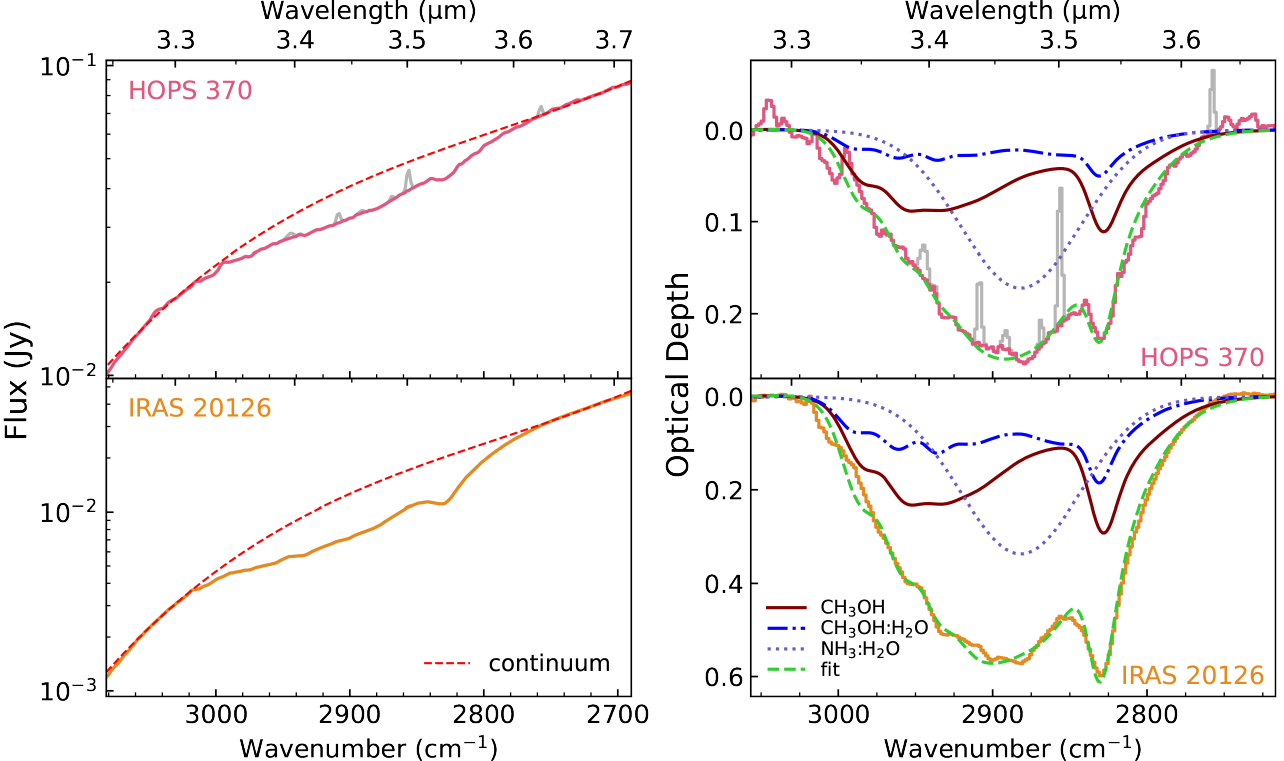}
\caption{Extraction and fitting of the CH$_{3}$OH 3.53 $\mu$m band. Left: adopted local continua. Right: fits to the extracted optical depth spectra. Some of the substructures present at 3.4 $\mu$m in the high S/N spectrum of IRAS 20126 are well reproduced by the more prominent CH$_{3}$OH overtones and C-H asymmetric stretching modes that emerge in H$_{2}$O-rich CH$_{3}$OH spectra.}
\label{fig:ch3oh_fit}
\end{figure*}

The full extracted spectra of both sources are presented in Figure~\ref{fig:all_spectra}. At first glance, it is apparent that in both spectra, the 3 $\mu$m O-H stretching modes of H$_{2}$O ice have sharp peaks at 3.1 $\mu$m, characteristic of a crystalline component \citep{hagen1981infrared,smith1989absorption}. This indicates that the ices along these lines of sight have experienced enough thermal processing to crystallize some of the water ice. The presence of pure CO$_{2}$ components in the $^{13}$CO$_{2}$ bands \citep{brunken2024jwst}, enhanced OCN$^{-}$ abundances \citep{nazari2024hunt}, and the relatively low absorptions of the CO ice features at 4.67 $\mu$m relative to the CO$_{2}$ features at 4.27 $\mu$m further corroborate that a significant portion of the ices toward these sources have experienced a substantial degree of thermal processing.

The ice species of interest to this work were analyzed as follows: first, the CH$_{3}$OH ice column densities and physicochemical environments were determined from the $\sim$3.53 $\mu$m ($\sim$2830 cm$^{-1}$) C-H stretching mode (Section~\ref{txt:ch3oh_ice}). This was done prior to the analysis of the HDO ice because CH$_{3}$OH has several weak combination modes between $\sim$3.7-4.2 $\mu$m ($\sim$2700-2400 cm$^{-1}$), one of which overlaps with the HDO O-D stretching mode at 4.1 $\mu$m; therefore, pre-constraining the expected contribution of CH$_{3}$OH ice to this region via a column density derived from a stronger band in a different region is necessary to ensure that any CH$_{3}$OH absorption at 4 $\mu$m is not misattributed to HDO ice (Section~\ref{txt:4um_region}), a concern previously raised by \cite{dartois2003revisiting}. The HDO ice column densities and morphologies were then determined via fitting the observed absorptions in the $\sim$3.7-4.2 $\mu$m region (Section~\ref{txt:hdo_ice}). Finally, the 3 $\mu$m ($\sim$3300 cm$^{-1}$) H$_{2}$O O-H stretching mode was fit to corroborate the morphologies found for HDO ice and to determine the ice HDO/H$_{2}$O ratios toward these sources (Section~\ref{txt:h2o_ice}).

\subsection{CH$_{3}$OH ice}
\label{txt:ch3oh_ice}

\begin{figure*}
\centering
\includegraphics[width=\linewidth]{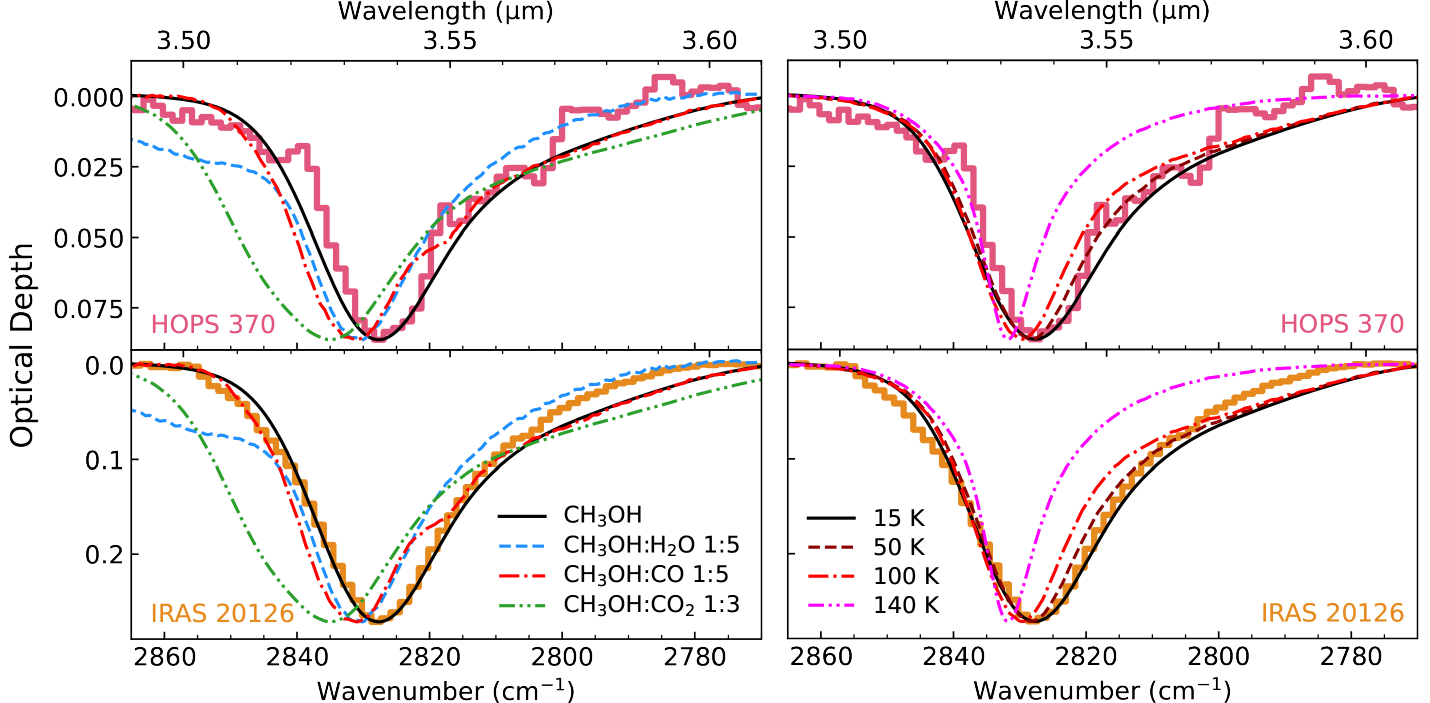}
\caption{The observed CH$_{3}$OH C-H symmetric stretching modes compared to select laboratory data. The peaks were extracted from the 3.5 $\mu$m absorption complexes via a local continuum fit using baseline points between $\sim$3.47-3.50 and 3.60-3.62 $\mu$m (Figure~\ref{fig:ch3oh_fit}). It is clear from these comparisons that cold, amorphous, and relatively pure CH$_{3}$OH is the major contributor to this feature. Left: observed peaks versus laboratory spectra of CH$_{3}$OH ice in various chemical environments. All spectra were collected at 15 K except the CH$_{3}$OH:CO$_{2}$ spectrum, which was collected at 10 K \citep{ehrenfreund1999laboratory}. Right: observed peaks versus laboratory spectra of pure CH$_{3}$OH ice at various temperatures.}
\label{fig:ch3oh_chem_temp}
\end{figure*}

Both sources present a clear CH$_{3}$OH C-H stretching mode at 3.53 $\mu$m. The CH$_{3}$OH ice column density can be obtained from this mode via a procedure outlined in \citet{brooke1999new} and \citet{boogert2022survey}. Briefly, this involves subtracting a local continuum using data points $\sim$3.25-3.3 and 3.6-3.7 $\mu$m and then fitting a Gaussian representing ammonia hydrates with a peak position of 3.47 $\mu$m (2881.844 cm$^{-1}$) and a full-width half maximum (FWHM) of 0.11 $\mu$m (91.4 cm$^{-1}$) along with laboratory CH$_{3}$OH ice spectra to the extracted feature. An identical local continuum subtraction is performed on the laboratory data as well prior to fitting the observed data to isolate the C-H stretching modes from the strong and broad O-H stretching feature with which they overlap. The resulting CH$_{3}$OH fits, along with the local continuum subtractions used to extract the CH$_{3}$OH features into the optical depth scale, are shown in Figure~\ref{fig:ch3oh_fit}. 

Toward both sources, the peak position of the 3.53 $\mu$m band is most consistent with laboratory data of cold ($<$60 K), amorphous, relatively pure CH$_{3}$OH ice (Figures~\ref{fig:ch3oh_chem_temp} and~\ref{fig:ch3oh_pos}). The 3.53 $\mu$m bands of laboratory CH$_{3}$OH ices that is warm, crystalline, or diluted with any of the three most abundant observable interstellar ice components (H$_{2}$O, CO, and CO$_{2}$) are too blue-shifted to fit the observed absorption. However, fitting the entire 3.5 $\mu$m region with only pure amorphous CH$_{3}$OH ice results in the spectrum being overfit at 3.4 $\mu$m by the blended CH$_{3}$OH overtones and C-H asymmetric stretching modes.

\begin{figure*}
\centering
\includegraphics[width=\linewidth]{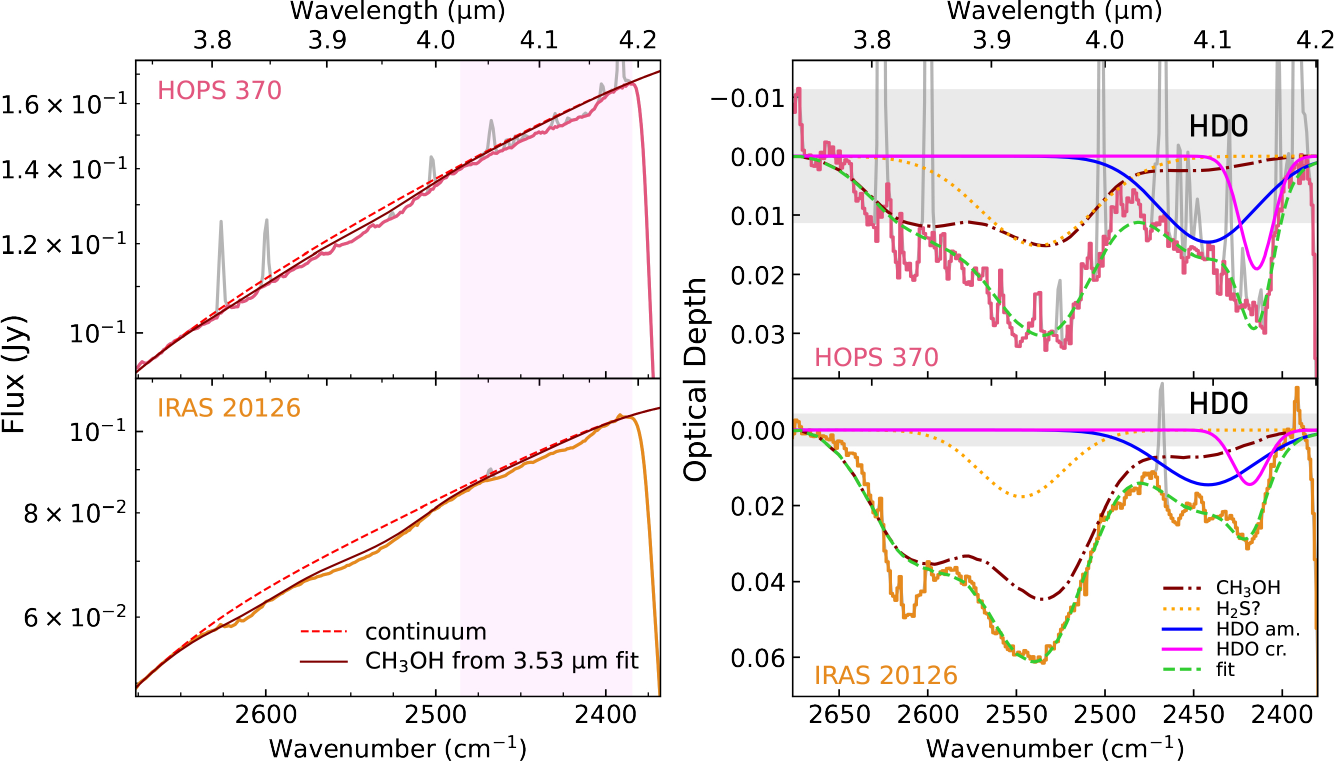}
\caption{Extraction and fitting of the 4 $\mu$m region. Left: adopted local continua, along with CH$_{3}$OH spectra determined from the 3.53 $\mu$m band fitting used to aid the continuum placement. The spectral region of the HDO O-D stretching mode is indicated via magenta shading. Right: fits to the extracted optical depth spectra (am. = amorphous; cr. = crystalline). The shaded gray regions indicate 3$\sigma$ uncertainty levels. The same combination of pure and H$_{2}$O-rich CH$_{3}$OH spectra as used to fit the 3.53$\mu$m feature is used in this fit. The best-fitting temperatures of the amorphous/crystalline HDO ice components are 132 K/47 K for HOPS 370 and 138 K/90 K for IRAS 20126, respectively (see Sections~\ref{txt:hdo_ice} and Appendix~\ref{app:chi2} for more details).}
\label{fig:hdo_fits}
\end{figure*}

This issue of pure CH$_{3}$OH ice spectra causing overfitting at 3.4 $\mu$m has previously been documented toward some LYSOs and MYSOs and was solved by using spectra of CH$_{3}$OH mixed with H$_{2}$O \citep{brooke1999new,pontoppidan2003detection}, which decreases the optical depth of the 3.4 $\mu$m features relative to that of the 3.53 $\mu$m feature. Such a decrease is not observed in CO- or CO$_{2}$-rich CH$_{3}$OH mixtures or in crystalline CH$_{3}$OH ices. Using a combination of two laboratory amorphous CH$_{3}$OH spectra, one of pure CH$_{3}$OH ice and an H$_{2}$O-rich CH$_{3}$OH ice mixture (1:5 CH$_{3}$OH:H$_{2}$O), in the fitting procedure results in satisfactory fits of both the 3.53 $\mu$m band and the 3.4 $\mu$m region (Figure~\ref{fig:ch3oh_fit}). A similar combination of pure CH$_{3}$OH and CH$_{3}$OH:H$_{2}$O laboratory spectra was used by \cite{dartois1999methanol} to fit the 3.53 and 3.9 $\mu$m features in the spectra observed toward MYSOs W33A and RAFGL 7009s.

The resulting CH$_{3}$OH column densities (reported in Table~\ref{tab:column_densities_ch3oh}) are obtained using the band strength of 4.86$\times$10$^{-18}$ cm molec$^{-1}$ for the C-H stretching mode of pure amorphous CH$_{3}$OH at 20 K \citep{luna2018densities}. This band strength was chosen because the profile of the 3.53 $\mu$m band indicates that the majority of the detected CH$_{3}$OH ice along these lines of sight is pure, amorphous, and cold ($>$60 K). The ratio of pure CH$_{3}$OH ice to CH$_{3}$OH ice mixed with H$_{2}$O that results in the best fits is highly dependent on the dilution factor of the CH$_{3}$OH:H$_{2}$O mixture used, so separate column densities for each of the two CH$_{3}$OH components were not calculated. Any effects of dilution with H$_{2}$O on the band strength of this feature are expected to be well within the 20\% uncertainty reported for the pure band strength \citep{kerkhof1999infrared,luna2018densities}. The reported uncertainties are explained in Appendix~\ref{app:cont_err_ch3oh}. The implications of the non-detection of warm or crystalline CH$_{3}$OH ice toward these lines of sight are further discussed in Section~\ref{txt:ch3oh_ice_formation}.

\begin{table}[h]
\caption{Derived CH$_{3}$OH ice column densities and abundances with respect to H$_{2}$O ice.}
\begin{center}
\begin{tabular}{c c c}
\hline
        Source & \multicolumn{2}{c}{CH$_{3}$OH} \\
        & (10$^{17}$ cm$^{-2}$) & (\% H$_{2}$O) \\
        \hline
        HOPS 370 & 7.8$\pm$1.6 & 16$\pm$6 \\
        IRAS 20126 & 22$\pm$4 & 27$\pm$10 \\
    \hline
     %   \noalign{\smallskip}
\label{tab:column_densities_ch3oh}
\end{tabular}
\end{center}
\end{table}

\subsection{4 $\mu$m region: CH$_{3}$OH, S-bearing species, and HDO}
\label{txt:4um_region}

A local continuum in the 3.7-4.2 $\mu$m (2700-2380 cm$^{-1}$) region was adopted so that the optical depth of the CH$_{3}$OH combination modes predicted by the CH$_{3}$OH column density derived from the 3.53 $\mu$m band did not exceed the optical depth of the continuum-subtracted spectrum, providing a constraint to prevent over-subtraction of features due to continuum choice in this region. The RMS error in optical depth was calculated for this region using data from 3.70-3.77 $\mu$m (2700-2650 cm$^{-1}$). It is important to note that, as can be seen in the right panel of Figure~\ref{fig:all_spectra}, the warping of the continuum by the scattering wings of the strong 4.27 $\mu$m CO$_{2}$ feature is very minimal in both spectra in comparison to some other icy lines of sight (e.g., \citealt{dartois2022influence,dartois2024spectroscopic}), greatly reducing the uncertainty of the local continuum placement in this region. The adopted local continua, the subtracted spectra in optical depth units, and the 3$\sigma$ uncertainty levels from the RMS error are presented in Figure~\ref{fig:hdo_fits}. 

Following this procedure, both lines of sight have excess absorption on top of the CH$_{3}$OH feature at $\sim$3.9 $\mu$m ($\sim$2550 cm$^{-1}$). \cite{jimenez2011sulfur} noted a similar excess absorption in the spectrum of the MYSO W33A and used it to derive an H$_{2}$S ice upper limit. However, \cite{hudson2018infrared} recently pointed out that many other S-H bond-containing ices, like CH$_{3}$SH and CH$_{3}$CH$_{2}$SH, have strong absorption features in this spectral region with nearly identical peak profiles, making it difficult to conclusively assign the broad excess absorption in this region to a specific molecule. Additionally, there is a lack of publicly available laboratory ice spectra of many of these sulfur-bearing species in astrophysically relevant mixtures. As sulfur-bearing ices are not the focus of this paper and the fitting of this feature has little effect on the analysis of the HDO ice feature, we simply fit this excess with a Gaussian and leave further analysis to a future work.

A small, narrow excess absorption is also present in both lines of sight at $\sim$3.83 $\mu$m ($\sim$2611 cm$^{-1}$), in addition to a sharp peak toward IRAS 20126 at $\sim$4.07 $\mu$m ($\sim$2459 cm$^{-1}$). We initially suspected these features could be attributed to crystalline CH$_{3}$OH ice, which has two sharp peaks at 3.83 and 4.07 $\mu$m (2612 and 2455 cm$^{-1}$) in the pure form. However, the entire CH$_{3}$OH combination mode complexes over the full 3.7-4.2 $\mu$m range could not be satisfactorily fit with any available crystalline CH$_{3}$OH ice spectra, pure or mixed. The observed spectra are missing an additional sharp and distinct feature at 3.94 $\mu$m (2536 cm$^{-1}$) that is present in the crystalline CH$_{3}$OH laboratory spectra. Annealed crystalline CH$_{3}$OH spectra also did not provide a good fit. Furthermore, the 3.53 $\mu$m feature lacks any significant spectral contribution from crystalline CH$_{3}$OH ice. Because no other ice spectra available to us provided a convincing match, we refrain from assigning these features, although we do not completely exclude crystalline CH$_{3}$OH as a candidate carrier.

Finally, in both lines of sight, excess absorption emerges from the continuum between $\sim$4.04-4.17 $\mu$m ($\sim$2475-2400 cm$^{-1}$) which cannot be accounted for by the CH$_{3}$OH combination mode without inducing unrealistic inflections in the local continua or overfitting the CH$_{3}$OH combination modes between 3.8-3.9 $\mu$m. There are no currently known S-H bond-bearing species whose absorptions peak at these wavelengths \citep{hudson2018infrared}, and the features present in crystalline CH$_{3}$OH ice spectra in this spectral region are all too blue-shifted from the observed peak positions at $\sim$4.14 $\mu$m. We therefore assign these features to HDO ice in various morphological states.

\subsection{HDO ice}
\label{txt:hdo_ice}
We performed fits to the local continuum-subtracted 4.04-4.17 $\mu$m region using a combination of CH$_{3}$OH ice laboratory spectra, a Gaussian representing S-H bond-bearing species, and HDO ice laboratory spectra. Prior to the fitting, the profiles of the HDO O-D stretching features were isolated from the underlying broad H$_{2}$O combination modes in the 0.4\% HDO:H$_{2}$O ice spectra (see Figure~\ref{fig:hdo_extraction}) and subsequently smoothed with Gaussian fits to eliminate experimental noise. The optical depths of the CH$_{3}$OH combination modes (as constrained by the CH$_{3}$OH column density derived from the 3.53 $\mu$m band in Section~\ref{txt:ch3oh_ice}) and the Gaussian fit to the 3.9 $\mu$m excess were not varied in the fitting procedure. The same combinations of pure and H$_{2}$O-rich CH$_{3}$OH ice spectra that were used to fit the 3.53 $\mu$m feature were also used here.

Toward both sources, a minimum of two HDO profiles, one amorphous and one crystalline, is needed to fit the HDO feature due to its asymmetric profile. The 3 $\mu$m H$_{2}$O bands toward both sources also have profiles characteristic of the lines of sight containing both amorphous and crystalline H$_{2}$O ice (Section~\ref{txt:h2o_ice}). This correlation between the detected presence of crystalline H$_{2}$O and crystalline HDO provides additional confidence in the peak assignment, because if some regions of the observed ice envelopes have been thermally processed enough to crystallize the H$_{2}$O ice, then any HDO ice in those regions should have crystallized as well.

\begin{table*}[h]
\caption{Derived HDO ice column densities and abundances with respect to H$_{2}$O ice.}
\begin{center}
\begin{tabular}{c c c c c}
\hline
        Source & Amorphous HDO & Crystalline HDO & \multicolumn{2}{c}{Total HDO} \\
        & (10$^{16}$ cm$^{-2}$) & (10$^{15}$ cm$^{-2}$) & (10$^{16}$ cm$^{-2}$) & (\% H$_{2}$O) \\
        \hline
        HOPS 370 & 1.6$\pm$0.8 & 6.5$\pm$0.5 & 2.3$\pm$0.8 & 0.46$\pm$0.22 \\
        IRAS 20126 & 1.6$\pm$0.9 & 5.0$\pm$0.9 & 2.1$\pm$0.9 & 0.26$\pm$0.14 \\
    \hline
     %   \noalign{\smallskip}
\label{tab:column_densities_hdo}
\end{tabular}
\end{center}
\end{table*}

\begin{figure}
\centering
\includegraphics[width=\linewidth]{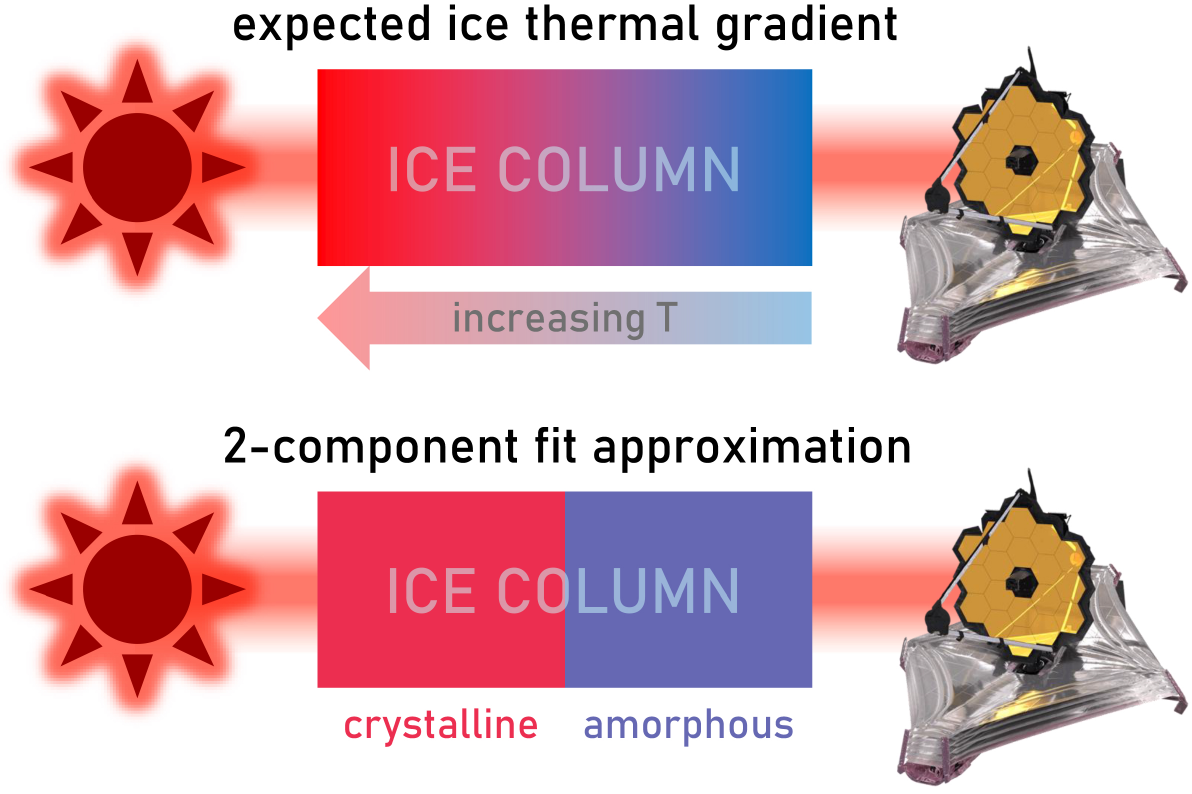}
\caption{A simplified diagram showing the difference between the expected thermal profile of the observed ice columns and the two-component fit approximation used to fit the observations in this work.}
\label{fig:ice_gradient}
\end{figure}

In total, 84 amorphous HDO ice spectra between 15-140 K and 22 crystalline HDO ice spectra between 15-150 K were collected. Every possible combination of one amorphous profile and one crystalline profile out of these spectra were fit to the observed spectra using a least-squares fitting procedure. The least-squares fits that result in the lowest $\chi^{2}$ value for both lines of sight are presented in Figure~\ref{fig:hdo_fits}. The relationships between the fit $\chi^{2}$ values and the temperatures of the laboratory spectra are shown in Figure~\ref{fig:chi2_plots}.

It is important to emphasize that the temperatures resulting in the lowest $\chi^{2}$ values should not be interpreted as the actual temperatures of the observed ices for four reasons. First, laboratory ice temperatures do not correspond directly to interstellar ice temperatures due to the faster heating rates and higher pressures experienced by laboratory ices \citep{redhead1962thermal,minissale2022thermal}. This means that, in the laboratory, ices crystallize and desorb at higher temperatures than in the interstellar medium. Second, these fits are not guaranteed to be unique, especially given that the broad, weak profile of the amorphous HDO could be considered degenerate at a wide range of temperatures when accounting for fitting errors caused by factors such as the observed spectral noise and uncertainties in the local continuum choice. This is particularly clear from Figure~\ref{fig:chi2_plots}, which shows that a wide range of HDO temperatures produce fits that are degenerate within 3$\sigma$. Third, it is possible that much of the HDO ice along these lines of sight is located in different chemical environments than just pure H$_{2}$O ice, which could further alter the HDO band profile. For example, models from \cite{taquet2014multilayer} and \cite{furuya2016reconstructing} suggest that the majority of HDO ice is co-produced with CH$_{3}$OH ice after the heavy CO freeze-out stage. Changes in the amorphous HDO feature caused by such chemical environments are difficult to unambiguously characterize in the laboratory due to the spectral overlap between the main HDO band and the CH$_{3}$OH combination modes. However, such chemical environments would not be expected to affect the crystalline component because CH$_{3}$OH and CO both desorb at temperatures lower than the water ice crystallization temperature, so by the time HDO ice crystallizes, it is expected to have been effectively distilled by heating. And finally, fitting two laboratory spectra at discrete temperatures to the observations is merely an approximation of the expected continuous thermal gradient experienced by the observed ices along each line of sight (Figure~\ref{fig:ice_gradient}).

Nonetheless, such an approximation is still useful for evaluating the thermal structure of the ice envelope qualitatively \citep{smith1989absorption}. In particular, the presence of a crystalline HDO component indicates that at least some regions within the ice envelopes must have experienced temperatures high enough to crystallize water ice, while the presence of an amorphous component indicates that other regions of the ice envelopes must have remained colder. Furthermore, the peak position of the crystalline HDO component, which suffers from much less uncertainties due to its sharper profile, matches best with annealed laboratory spectra, hinting at a possibility of heated ices being re-cooled within the ice envelope (see further discussion in Section~\ref{txt:thermal_cycling}).

The uncertainty in the temperature of the observed ices translates to an uncertainty in any HDO ice column density calculations because the apparent band strength of the O-D stretching feature in HDO measured in the laboratory varies with temperature, particularly for the amorphous phase (see Figure~\ref{fig:hdo_band_strength}). Between 15 and 135 K, the amorphous HDO apparent band strength increases by a factor $\sim$1.7 from 4.2$\times$10$^{-17}$ to 7.1$\times$10$^{-17}$ cm molec$^{-1}$. This uncertainty is much smaller for the crystalline HDO component because its apparent band strength only varies by $\sim$8\% between 15-150 K. When calculating the HDO column densities from the fits, we used band strengths of 6.0$\times$10$^{-17}$ and 6.7$\times$10$^{-17}$ cm molec$^{-1}$ for the amorphous and crystalline HDO ice features, respectively, which correspond to intermediate laboratory ice temperatures of around 80 K for the amorphous component and 90 K for the crystalline component (see Table~\ref{tab:hdo_peak_char}). The resulting HDO ice column densities are listed in Table~\ref{tab:column_densities_hdo}, and the reported uncertainties are explained in Appendix~\ref{app:cont_err_hdo}.

\begin{figure*}[h]
\centering
\includegraphics[width=\linewidth]{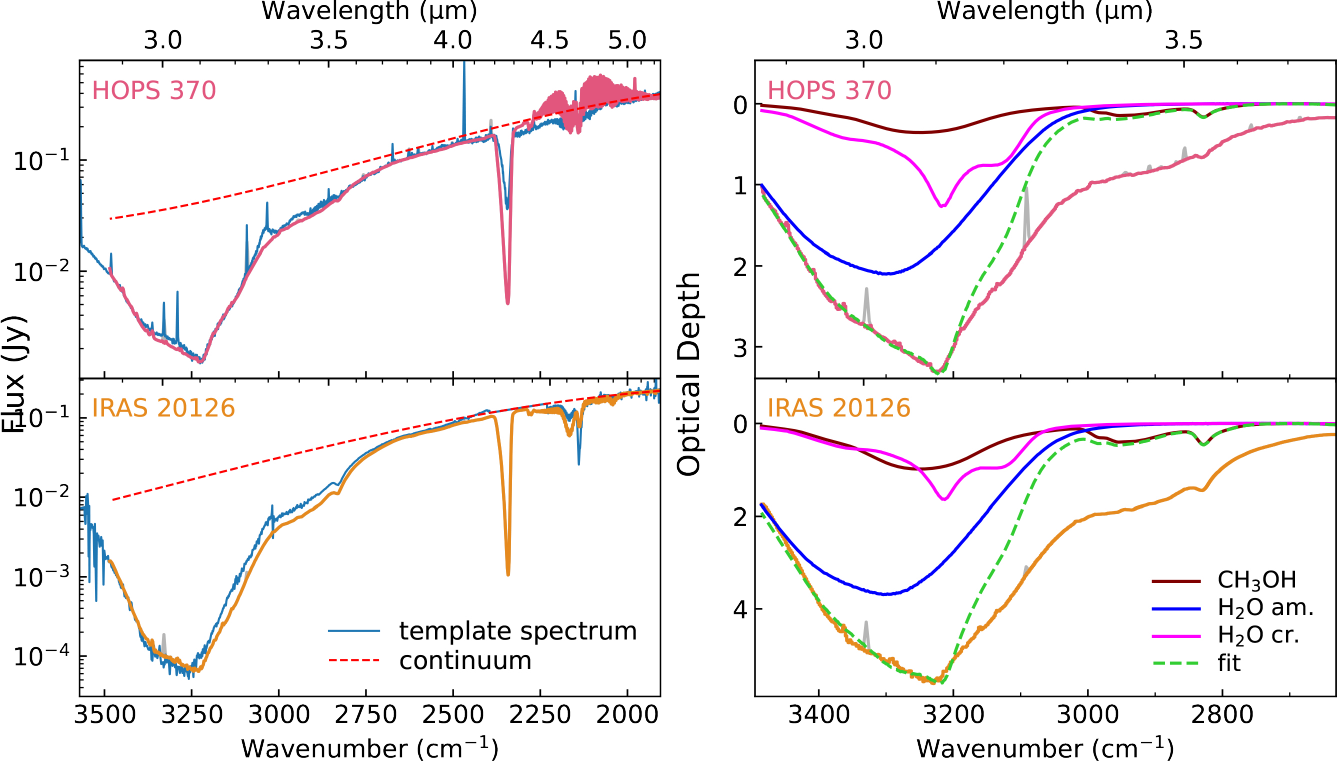}
\caption{Extraction and fitting of the H$_{2}$O 3 $\mu$m band. Left: adopted global continua and the archival MYSO ice spectra used as templates to guide the continuum choice. Right: fits to the extracted optical depth spectra (am. = amorphous; cr. = crystalline).}
\label{fig:h2o_fits}
\end{figure*}

Within the uncertainties, the ratio of amorphous:crystalline HDO ice column densities are calculated to be between $\sim$1.2-4 for HOPS 370 and between $\sim$1.2-6 for IRAS 20126 (i.e., there is more amorphous than crystalline HDO ice toward both sources). However, it must be emphasized once more that the uncertainties of the assignment, fits, and calculated column densities are greater for the amorphous HDO component because the amorphous profile is significantly weaker and broader and its apparent band strengths vary to a much greater extent with temperature, so more precise ratios of amorphous:crystalline HDO ice cannot be determined.

% a range of initial guess params used to ensure best fit 

\begin{table*}[h]
\caption{Derived H$_{2}$O ice column densities.}
\begin{center}
\begin{tabular}{c c c c}
\hline
        Source & Amorphous H$_{2}$O & Crystalline H$_{2}$O & Total H$_{2}$O \\
        & (10$^{18}$ cm$^{-2}$) & (10$^{17}$ cm$^{-2}$) & (10$^{18}$ cm$^{-2}$) \\
        \hline
        HOPS 370 & 4.0$\pm$1.7 & 9.9$\pm$3.1 & 5.0$\pm$1.7 \\
        IRAS 20126 & 7.0$\pm$2.7 & 12$\pm$4.1 & 8.3$\pm$2.8 \\
    \hline
     %   \noalign{\smallskip}
\label{tab:column_densities_h2o}
\end{tabular}
\end{center}
\end{table*}

\subsection{H$_{2}$O ice}
\label{txt:h2o_ice}

To estimate the ice HDO/H$_{2}$O ratios along these lines of sight, we quantified the H$_{2}$O ice column density from the 3 $\mu$m band by defining global continua over the full NIRSpec G395M range and subsequently fitting the 3 $\mu$m band with laboratory H$_{2}$O ice spectra. The global continua and least-squares fits of the 3 $\mu$m feature are presented in Figure~\ref{fig:h2o_fits}.

An important caveat in the analysis of this feature is the lack of collected continuum data points below 2.9 $\mu$m, resulting in greater uncertainties in the slope of the blue wing and optical depth of the 3 $\mu$m feature in the global continuum-subtracted spectra. We attempted to mitigate this issue by using spectra of other MYSO ice envelopes as templates to guide our global continuum choice. These spectra contain continuum data points $<$2.9 $\mu$m, making their global continuum determinations and subsequent H$_{2}$O profile extractions much more reliable, and their H$_{2}$O profiles are similar to those in the spectra investigated in this work. The best-matching H$_{2}$O profiles found for HOPS 370 and IRAS 20126 were those of Orion from ISO (shown in \citealt{dartois2001search}) and G034.7123-00.5946 from IRTF (shown in \citealt{boogert2022survey}), respectively. The continuum points on the blue side of the 3 $\mu$m band were then defined so that the profile of the 3 $\mu$m feature in the JWST spectra replicated the profile in the template spectra as closely as possible (see Figure~\ref{fig:h2o_fits}).

As done in previous studies \citep{smith1989absorption,dartois2001search}, the H$_{2}$O band in the optical depth scale was then fit with multiple components as an approximation to the thermal gradient expected along the line of sight. The fitting procedure was analogous to that performed on the 4 $\mu$m region in that the best-fitting combination of one amorphous and one crystalline H$_{2}$O lab spectrum was determined by calculating the $\chi^{2}$ values of the least-squares fits of every possible combination. As stressed in Section~\ref{txt:hdo_ice}, the temperatures of the lab H$_{2}$O ices fit to the observations should not be taken as indications of the real physical temperature of the observed ices, as these two-component fits are only simple approximations of the thermal gradient that is expected along these lines of sight, and they are not guaranteed to be unique; however, it is certain that some combination of cold and warm ices is needed to sufficiently fit the feature toward both sources. Furthermore, similarly to the analysis performed by the ICE AGE team in \citet{mcclure2023ice}, grain shape effects were not modeled in these fits, as a detailed characterization of the H$_{2}$O 3 $\mu$m band is outside our primary scope (quantifying H$_{2}$O to obtain HDO/H$_{2}$O ratios). Indeed, recently published follow-up analyses of grain shape effects in the ICE AGE spectra show that their derived relative ice abundances are largely the same regardless of whether their spectral fitting procedure includes grain shape effects \citep{dartois2024spectroscopic}.

The contribution of the O-H stretching mode of CH$_{3}$OH ice to the observed 3 $\mu$m feature was also approximated similarly to the HDO fitting procedure, where the column density of CH$_{3}$OH was set as a constant value from the fits to the 3.53 $\mu$m band. However, a spectrum of only pure CH$_{3}$OH was used to model the contribution of CH$_{3}$OH to the 3 $\mu$m feature, as it is difficult to accurately deconvolve the overlapping O-H stretching modes of CH$_{3}$OH and H$_{2}$O in the CH$_{3}$OH:H$_{2}$O lab spectra. Contributions to the red wing by ammonia hydrates were neglected. For this reason (as well as the neglect of grain shape effects), only the region from 2.88-3.13 $\mu$m (3470-3191 cm$^{-1}$) was considered in the fitting procedure, and the long-wavelength wing of the 3 $\mu$m feature remains underfit. This underfitting is taken into account in our reported uncertainties (see Appendix~\ref{app:cont_err_h2o}).

Similar to the band strength of the HDO O-D stretch, the band strength of the H$_{2}$O O-H stretch varies with temperature. The reported magnitude of this variation differs in the literature \citep{hagen1981infrared,gerakines1995infrared,mastrapa2009optical}. In our experiments, the integrated absorbance of the O-H stretch increases by a factor of up to 1.26 between 15 and 135 K in amorphous H$_{2}$O, while there is only a factor of 1.16 difference between the integrated absorbance at 150 and 15 K in crystalline H$_{2}$O. These variations are again treated as a source of uncertainty in the reported H$_{2}$O column densities (see Appendix~\ref{app:cont_err_h2o}).

To quantify H$_{2}$O, all of the fit H$_{2}$O components were integrated, and a band strength of 1.9$\times$10$^{-16}$ cm molec$^{-1}$ \citep{mastrapa2009optical} was used for the amorphous component. For the crystalline component, a band strength of 2.5$\times$10$^{-16}$ cm molec$^{-1}$ was used, which is the apparent band strength we calculated for the crystalline H$_{2}$O O-H stretch below 120 K in our laboratory spectra. The resulting values are reported in Table~\ref{tab:column_densities_h2o}. Within the error margins, the ratios of amorphous:crystalline H$_{2}$O ice column densities range from 3.8-5.8 toward HOPS 370 and 5.7-9.1 toward IRAS 20126. These amorphous:crystalline ratios are higher than those of HDO but remain in agreement when considering uncertainties. We refrain from drawing chemical conclusions from the comparison between amorphous and crystalline column densities of HDO and H$_{2}$O ice for reasons discussed in Section~\ref{txt:inferred_ratios}.

\section{Discussion}

\subsection{Inferred ice HDO/H$_{2}$O ratios}
\label{txt:inferred_ratios}

\begin{figure*}[ht!]
\centering
\includegraphics[width=\linewidth]{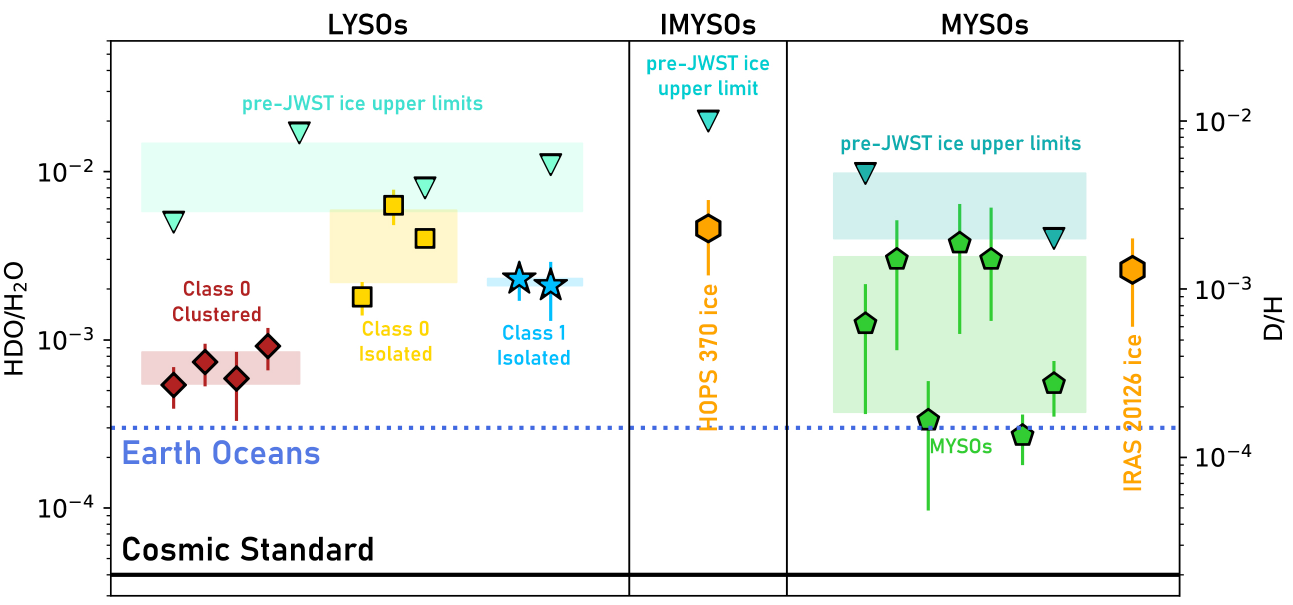}
\caption{Comparison of HDO/H$_{2}$O ratios measured in the gas and ice toward protostars of various masses (adapted from \citealt{jensen2019alma} and \citealt{andreu2023high}). Ratios measured in the gas phase are plotted with red diamonds for clustered Class 0 LYSOs, yellow squares for isolated Class 0 LYSOs, blue stars for isolated Class 1 LYSOs, and green pentagons for MYSOs. Ice upper limits derived from pre-JWST spectra are plotted with teal triangles. The JWST ice values from this work are plotted with orange hexagons. The literature values used in this figure are provided in Table~\ref{tab:hdo_h2o_lit}.}
\label{fig:gas_vs_ice}
\end{figure*}

The ice HDO/H$_{2}$O ratios measured toward HOPS 370 and IRAS 20126, 4.6$\pm$2.2$\times$10$^{-3}$ and 2.6$\pm$1.4$\times$10$^{-3}$, respectively, are both two orders of magnitude greater than what is expected from the cosmic standard D/H abundance. This enhancement far exceeds our reported error margins and is consistent with the deuterated water ice that we detect along these lines of sight having cold, prestellar origins. The ice HDO/H$_{2}$O of the IMYSO is higher by almost a factor of 2 compared to the MYSO, but this difference is not significant within the error margins of both values. A larger sample size of ice detections in IMYSOs and MYSOs is needed to draw concrete conclusions regarding any systematic differences between IMYSO and MYSO ice HDO/H$_{2}$O ratios.

While the ice HDO/H$_{2}$O measured toward these protostars are expected to largely reflect their prestellar conditions and timescales, experimental works have shown that it is also possible for the prestellar D/H ratios of protostellar ices to be altered prior to sublimation via inter-species deuterium exchange reactions during ice thermal processing. Specifically, deuterium exchange reactions proceed very efficiently in warm ices ($T$$\gtrsim$70-120 K) between atoms involved in hydrogen bonds in a variety of astrophysically relevant simple molecules \citep{kawanowa2004hydration,ratajczak2009hydrogen,faure2015kinetics,lamberts2015thermal} and are particularly promoted between water molecules following the phase transition from amorphous to crystalline \citep{galvez2011hdo}.

It may therefore be tempting to compare the ice HDO/H$_{2}$O ratios of the separate amorphous and crystalline components to investigate if a difference in deuteration ratio in the unprocessed versus thermally processed ices can be observed. Although the HDO/H$_{2}$O ice ratios of the reported crystalline components are higher than those of the amorphous components toward both sources (i.e., the ratios of amorphous:crystalline column densities are lower for HDO than for H$_{2}$O), we caution against drawing chemical conclusions from these differences for multiple reasons: 1) the differences are close to the reported error margins and therefore are far from statistically significant; 2) because the two-component fits only approximate the expected thermal gradients, the relative abundances of amorphous to crystalline ices are also only approximate; and 3) grain shape effects were not taken into account when the H$_{2}$O bands were fit. This last point is particularly important to remember because, while including grain shape effects in spectral fitting procedures may not greatly affect the derived overall relative molecular abundances of ice species, the warping of the strongest ice bands by grain shape effects do preclude detailed quantitative analysis of ice temperature, morphology, or chemical environment via the profiles of such bands without accounting for grain shape correction \citep{dartois2024spectroscopic}.

\subsection{Comparing ice and gas HDO/H$_{2}$O ratios}
Figure~\ref{fig:gas_vs_ice} provides a comparison of the ice HDO/H$_{2}$O ratios measured in this work with previously published protostellar gas phase ratios and ice upper limits from predecessor IR telescopes.

The MYSO gas HDO/H$_{2}$O ratios reported in the literature over the past couple decades span an order of magnitude from values as low as 2-3$\times$10$^{-4}$ \citep{emprechtinger2013abundance,coutens2014water} to as high as 2-3$\times$10$^{-3}$ \citep{van2006water,neill2013abundance}. The ice HDO/H$_{2}$O ratio measured in this work toward the MYSO IRAS 20126 is in agreement with the highest gas-phase MYSO HDO/H$_{2}$O ratios. The agreement supports the frequently made assumption that deuterium abundances detected in the hot inner regions of protostars are representative of deuterium abundances in ices.

An ice HDO/H$_{2}$O ratio on the order of 10$^{-3}$ toward a MYSO is also consistent with the conclusions drawn by \cite{van2006water}, who observed an inverse correlation between the gas HDO/H$_{2}$O ratios and the mass-weighted average envelope temperatures of the four MYSOs they studied. They interpreted this as evidence that the HDO/H$_{2}$O ratios on the order of 10$^{-3}$ (measured toward the coldest envelopes) were the most representative of ice HDO/H$_{2}$O ratios, while the HDO/H$_{2}$O ratios on the order of 10$^{-4}$ (measured toward the warmer, more evolved envelopes) had already been partially altered by gas-phase reactions in the hot cores where water ice desorbs. It is interesting to also note that, according to the fits of \cite{boogert2000iso}, the $^{13}$CO$_{2}$ ice features of the two sources studied by \cite{van2006water} with lower HDO/H$_{2}$O ratios show a significant degree of CO$_{2}$ ice segregation and therefore thermal processing, while the $^{13}$CO$_{2}$ ice feature of the \cite{van2006water} source in which the gas HDO/H$_{2}$O ratio was the highest, W33A, is indicative of a much lower degree of thermal processing.

However, it is also possible that MYSOs could vary significantly in their prestellar timescales and conditions, and therefore their ice HDO/H$_{2}$O ratios may greatly differ. A larger sample size of MYSO ice HDO/H$_{2}$O measurements will be needed to ascertain what causes the observed spread in MYSO gas HDO/H$_{2}$O values. In any case, the agreement between the IRAS 20126 ice HDO/H$_{2}$O ratio and the gas HDO/H$_{2}$O ratios of several other MYSOs suggests that the water gas detected toward at least some hot cores has not been substantially processed by gas-phase reactions.

\citet{van2006water} additionally noted that the gas HDO/H$_{2}$O ratios measured toward their MYSOs are one to two orders of magnitude lower than those measured in LYSOs, consistent with the theory that LYSOs experience longer prestellar dense stages than MYSOs. An analogous trend is observed in the deuterium ratios of protostellar gas-phase CH$_{3}$OH, which are on average more than an order of magnitude lower in MYSOs than LYSOs; such a difference can be explained with models if MYSOs have shorter and/or warmer prestellar stages relative to LYSOs \citep{van2022methanol}. However, the most recent gas HDO/H$_{2}$O LYSO values that utilize interferometric techniques to target only the hot inner regions of these objects and detect multiple high S/N HDO lines indicate that some of the previously measured LYSO gas HDO/H$_{2}$O ratios were likely overestimated by one to two orders of magnitude \citep{persson2013warm}. In fact, recent gas phase ratios of isolated LYSOs (which are thought to experience long and cold dense stages) ranging from 1.8-6.3$\times$10$^{-3}$ \citep{jensen2019alma,jensen2021alma} are very similar to the highest measured MYSO gas ratios as well as the IMYSO and MYSO ice ratios measured here. The clustered LYSO gas ratios are on average a factor of a few lower, ranging from 5.4-9.2$\times$10$^{-4}$ \citep{persson2014deuterium,jensen2019alma}, and are more similar to the low-end of measured MYSO ratios.

This similarity in the recently published gas HDO/H$_{2}$O ratios of LYSOs and MYSOs could be interpreted in a number of ways. One potential explanation is that the most recently published gas-phase LYSO HDO/H$_{2}$O ratios have been altered by gas-phase reactions, meaning that the measured hot inner gas molecular abundances are not fully representative of the molecular abundances of icy species (similar to the interpretation of MYSO gas HDO/H$_{2}$O ratios on the order of 10$^{-4}$ made by \citealt{van2006water}). Measurements of ice HDO/H$_{2}$O ratios toward LYSOs may prove useful in resolving if this is the case.

Alternatively, some MYSOs may experience prestellar stages with timescales and temperatures similar to LYSOs. Measuring the gas-phase D$_{2}$O/HDO ratios in MYSOs could elucidate the credibility of this latter scenario. This is because D$_{2}$O/HDO ratios, especially when compared to HDO/H$_{2}$O ratios, are also an excellent diagnostic of formation conditions \citep{furuya2017water}. Multiple studies have measured gas D$_{2}$O/HDO ratios in LYSOs that are more than a factor of two higher than their gas HDO/H$_{2}$O ratios \citep{coutens2014high,jensen2021alma}, a result that can only be reproduced by models if the majority of deuterated water ice is produced in the cold dense prestellar stage after the heavy CO freeze-out stage alongside CH$_{3}$OH ice \citep{furuya2016reconstructing}.

Due to the extremely low abundance of D$_{2}$O, few gas D$_{2}$O/HDO ratios have been measured in either LYSOs or MYSOs. Currently, one D$_{2}$O/HDO ratio has been measured toward the hot core of the Orion KL MYSO \citep{neill2013abundance}. This value, 1.6$\times$10$^{-3}$, is an order of magnitude lower than the gas D$_{2}$O/HDO ratios that have been measured in LYSOs \citep{jensen2021alma}, which may indicate that less of the deuterated water detected toward Orion KL was produced in the cold dense prestellar stages. However, this is again a comparison of very few data points, and more D$_{2}$O/HDO measurements in both LYSOs and MYSOs are needed to draw such a conclusion with certainty.

\subsection{Ice envelope evolution}
\label{txt:env_structure}

In addition to the ice HDO/H$_{2}$O ratios, the ice fitting performed in this work can also provide information about the thermal and chemical evolution of the ice envelopes surrounding the investigated objects.

\subsubsection{Ice thermal cycling}
\label{txt:thermal_cycling}

The peak position of the crystalline components observed toward HOPS 370 and IRAS 20126 match best with that of annealed HDO ice (\textit{i.e.}, HDO that has been heated to its crystallization temperature and then cooled back down), as fits using non-annealed crystalline HDO ice at 150 K result in $\Delta\chi^{2}$ values that lie outside the 3$\sigma$ confidence intervals (Figure~\ref{fig:chi2_plots}). Such an observed red-shift from the laboratory crystalline HDO ice peak positions at 150 K may be indicative of thermal cycling within the protostellar envelopes, where ice that was heated to high enough temperatures to transition from an amorphous to a crystalline state was either transported back to cooler regions in the cold envelope (e.g., via motions driven by outflows), or the thermal gradient within the cold envelope changed (e.g., due to an accretion burst).

\subsubsection{CH$_{3}$OH ice formation}
\label{txt:ch3oh_ice_formation}

The CH$_{3}$OH ice abundances of 16\% and 27\% with respect to H$_{2}$O ice toward both sources are on the high end of the values typically measured for both low-mass and massive protostars (\citealt{boogert2015observations} and references therein). Although unusual, it is not unprecedented for such high values to be measured toward MYSOs \citep{dartois1999methanol}. These high CH$_{3}$OH ice abundances may be due to particularly long cold dense prestellar stages relative to the length of the warmer and less dense stages when the bulk of the H$_{2}$O ice is produced. As models suggest that the majority of HDO ice is produced during the cold dense prestellar stage with CH$_{3}$OH ice \citep{furuya2016reconstructing}, this would imply that the ice HDO/H$_{2}$O ratios measured here could also be on the high end of most protostars, and other protostars with lower CH$_{3}$OH ice abundances relative to H$_{2}$O ice that had shorter cold dense prestellar stages could then have lower ice HDO/H$_{2}$O ratios. However, it is also possible that thermal processing could affect CH$_{3}$OH ice abundances. Correlations between ice HDO/H$_{2}$O ratios, CH$_{3}$OH ice abundances, and the profiles of ice thermal processing tracers like $^{13}$CO$_{2}$ may be interesting to investigate once the sample size of measured ice HDO/H$_{2}$O ratios is larger.

The fact that the current fitting procedure of the 3.5 $\mu$m absorption complex requires a component of CH$_{3}$OH in H$_{2}$O-rich mixtures also carries astrochemical significance. Although the earliest fits to CH$_{3}$OH ice features toward protostars often utilized H$_{2}$O-rich CH$_{3}$OH ice mixtures \citep{allamandola1992infrared,dartois1999methanol,brooke1999new,pontoppidan2003detection}, subsequent studies popularized an ice evolution model where most of the observed CH$_{3}$OH ice forms after CO freeze-out in a relatively CO-rich and H$_{2}$O-poor ice environment. This model is supported by analysis of the profiles of the 9.75 $\mu$m CH$_{3}$OH ice C-O stretching mode \citep{bottinelli2010c2d}, the "red component" of the 4.67 $\mu$m CO ice C$\equiv$O stretching mode \citep{cuppen2011co,penteado2015spectroscopic}, the relationships between CH$_{3}$OH ice column densities, CO ice column densities, and visual extinction A$_{v}$ \citep{boogert2015observations}, and the experimentally demonstrated efficiency of solid-state CH$_{3}$OH formation via CO hydrogenation \citep{watanabe2002efficient,fuchs2009hydrogenation} 

However, recent experimental works indicate that CH$_{3}$OH can also be produced in H$_{2}$O-rich ices via a reaction between CH$_{4}$ and OH radicals \citep{qasim2018formation}, albeit less efficiently than via CO hydrogenation. Recent JWST data toward prestellar cores, where the 9.75 $\mu$m CH$_{3}$OH ice feature is well-fit with a combination of CO-rich and H$_{2}$O-rich CH$_{3}$OH ice mixtures, provide observational evidence supporting this pathway as a secondary means of prestellar CH$_{3}$OH ice production \citep{mcclure2023ice}. In our fits, the lack of a CO-rich CH$_{3}$OH ice component is likely due to the fact that most of the CO ice toward these lines of sight has thermally desorbed, as indicated by the low CO ice column density and presence of CO gas lines toward both lines of sight \citep{federman2023investigating,rubinstein2023ipa}, leaving behind a layer of nearly pure CH$_{3}$OH ice that likely formed via CO hydrogenation, as well as some CH$_{3}$OH ice in a water-rich chemical environment that may have formed during the less dense prestellar stages, perhaps via the CH$_{4}$ + OH pathway.

Notably, despite the presence of thermal processing tracers in both spectra, there is a lack of strong spectroscopic evidence of heated or crystalline CH$_{3}$OH ice along either line of sight. Perhaps this could be explained by the heated ices having mostly experienced temperatures too hot for CH$_{3}$OH to remain in the ice (i.e., a steep thermal gradient along the line of sight). However, both the dominant component of the CH$_{3}$OH feature being pure CH$_{3}$OH ice as well as the low optical depth of the CO ice feature indicate that the majority of the "cold" ices in the outermost envelope must have also experienced some amount of heating (equivalent to at least 30-40 K in the laboratory).

\section{Conclusions}

Here we present $>$3$\sigma$ detections of both amorphous and crystalline HDO ice at 4.1 $\mu$m in JWST NIRSpec spectra taken toward the IMYSO HOPS 370 and the MYSO IRAS 20126+4104. The HDO ice column densities were quantified in a fitting procedure that accounted for potential spectral contributions from CH$_{3}$OH ice. The H$_{2}$O ice column densities were quantified via the 3 $\mu$m feature to enable comparisons between HDO/H$_{2}$O ratios measured in the ice with those measured in the gas toward protostars. Our main findings are summarized as follows:

\begin{enumerate}

\item Toward both sources, the detection of a crystalline HDO ice component at 4.1 $\mu$m correlates with the detection of a crystalline H$_{2}$O ice component at 3 $\mu$m as well as other tracers of ice thermal processing, increasing confidence in our assignment of the 4.1 $\mu$m feature to HDO ice.

\item We measure ice HDO/H$_{2}$O ratios of 4.6$\pm$2.2$\times$10$^{-3}$ and 2.6$\pm$1.4$\times$10$^{-3}$ toward HOPS 370 and IRAS 20126, respectively. These ratios are enriched by approximately two orders of magnitude relative to the cosmic standard deuterium abundance, consistent with the water ice toward these objects having cold, prestellar origins.

\item The ice HDO/H$_{2}$O ratio measured toward the MYSO IRAS 20126 is consistent with the highest reported gas HDO/H$_{2}$O ratios toward MYSOs (a couple $\times$10$^{-3}$), supporting the assumption that the deuterium abundances of the gas-phase water observed toward these objects are representative of the deuterium abundances of their surrounding water ice.

\item The ice HDO/H$_{2}$O ratios toward these objects are remarkably similar to each other as well as to the gas HDO/H$_{2}$O ratios measured toward isolated LYSOs. A larger sample of ice HDO/H$_{2}$O ratios, both toward LYSOs and MYSOs, as well as more measurements of D$_{2}$O/H$_{2}$O ratios in the gas phase could help to explain the cause of this similarity.

\item The peak position of the crystalline HDO ice component toward both sources suggests that the thermally processed ice in these lines of sight experienced re-cooling to some degree.

\item The CH$_{3}$OH ice requires a water-rich CH$_{3}$OH ice mixture to obtain a satisfactory fit to the 3.4 $\mu$m region, supporting recent experimental and observational evidence that some CH$_{3}$OH ice may form prior to the heavy CO freeze-out stage in an H$_{2}$O-rich environment.

\item As part of this work, several new laboratory spectra of HDO, H$_{2}$O, CH$_{3}$OH, and CO ice mixtures are now available for public use on the Leiden Ice Database (LIDA).

\end{enumerate}

\begin{acknowledgements}
KS thanks Brian Ferrari, Adwin Boogert, Jenny Noble, Helen Fraser, Melissa McClure, Sergio Ioppolo, Neal Evans, and Alessio Caratti o Garatti for helpful discussions, Adwin Boogert for providing IRTF spectra, and Julia Santos for experimental support. This work is based on observations made with the NASA/ESA/CSA James Webb Space Telescope. The data were obtained from the Mikulski Archive for Space Telescopes at the Space Telescope Science Institute, which is operated by the Association of Universities for Research in Astronomy, Inc., under NASA contract NAS 5-03127 for JWST. These observations are associated with program \#1802. All the JWST data used in this paper can be found in MAST: \href{https://archive.stsci.edu/doi/resolve/resolve.html?doi=10.17909/3kky-t040}{10.17909/3kky-t040}. Astrochemistry at Leiden is supported by funding from the European Research Council (ERC) under the European Union’s Horizon 2020 research and innovation programme (grant agreement No. 101019751 MOLDISK), the Netherlands Research School for Astronomy (NOVA), and the Danish National Research Foundation through the Center of Excellence “InterCat” (Grant agreement no.: DNRF150). Support for SF, AER, STM, RG, JJT and DW in program \#1802 was provided by NASA through a grant from the Space Telescope Science Institute, which is operated by the Association of Universities for Research in Astronomy, Inc., under NASA contract NAS 5-03127. The National Radio Astronomy Observatory is a facility of the National Science Foundation operated under cooperative agreement by Associated Universities, Inc. Y.-L.Y. acknowledges support from Grant-in-Aid from the Ministry of Education, Culture, Sports, Science, and Technology of Japan (20H05845, 20H05844, 22K20389), and a pioneering project in RIKEN (Evolution of Matter in the Universe).

\end{acknowledgements}

%-------------------------------------------------------------------
\bibliographystyle{aa}
\bibliography{biblio}

\hrule
   {\tiny\noindent 1. Laboratory for Astrophysics, Leiden Observatory, Leiden University, P.O. Box 9513, 2300 RA Leiden, NL\\
   \email{slavicinska@strw.leidenuniv.nl}\\
   2. Leiden Observatory, Leiden University, P.O. Box 9513, NL 2300 RA Leiden, NL\\
   3. Max Planck Institut f\"ur Extraterrestrische Physik (MPE), Giessenbachstrasse 1, 85748 Garching, DE\\
   4. European Southern Observatory, Garching, DE\\
   5. Department of Physics and Astronomy, Bausch \& Lomb Hall, University of Rochester, Rochester, NY 14627, USA\\
   6. Department of Astronomy, University of Massachusetts Amherst, 710 North Pleasant Street, Amherst, MA 01003, USA\\
   7. Department of Astronomy \& Astrophysics Tata Institute of Fundamental Research, Homi Bhabha Rd, Colaba, Mumbai, Maharashtra, IN\\
   8. Academia Sinica Institute of Astronomy \& Astrophysics, No. 1 Sec. 4 Roosevelt Rd., Taipei 10617, TW, ROC\\
   9. Ritter Astrophysical Research Center, Department of Physics and Astronomy, University of Toledo, Toledo, OH 43606, USA\\
   10. Star and Planet Formation Laboratory, RIKEN Cluster for Pioneering Research, Wako, Saitama 351-0198, JP\\
   11. Department of Astronomy, University of Illinois, 1002 West Green St, Urbana, IL 61801, USA\\
   12. National Radio Astronomy Observatory, 520 Edgemont Rd., Charlottesville, VA 22903, USA\\
   13. Max Planck Institute for Astronomy, Heidelberg, Baden-W\"urttemberg, DE\\
   14. SKA Observatory, Jodrell Bank, Lower Withington, Macclesfield SK11 9FT, UK\\
   15. Department of Astronomy, University of Texas at Austin, 2515 Speedway, Stop C1400, Austin, TX 78712, USA\\
   16. INAF Osservatorio Astronomico di Capodimonte, Salita Moiariello 16, I-80131 Napoli, IT\\
   17. Friedrich-Schiller-Universit\"at, Jena, Th\"uringen, DE\\}

\begin{appendix}

\section{Laboratory data}
\label{app:lab}

Here we present the methodologies used to collect and analyze the laboratory data in this work.

\subsection{Experimental details}
In this work, independent dosing lines were used to deposit the CH$_{3}$OH-containing binary ice mixtures using the procedure outlined in \cite{yarnall2022new}, where each leak valve is calibrated so that its position results in the desired deposition rate of each mixture component in its pure form. After this calibration step, shut valves between the gas or vapor reservoirs and the leak valves are opened so that the mixture components deposit simultaneously upon the substrate. As mentioned in \cite{yarnall2022new}, the leak valve calibration depends on knowing the refractive index and density of each ice component, as well as the measurement of the laser interference pattern as the ice deposits in the calibration step. The benefits of utilizing such a system of independent dosing lines to create mixtures \textit{in situ} on the ice substrate rather than using a single dosing line and mixing gases and vapors prior to their introduction into the chamber have been previously discussed in \cite{gerakines1995infrared}, \cite{yarnall2022new}, and \cite{slavicinska2023hunt}.

A HeNe laser with a wavelength of 632.8 nm was used for the laser interference measurements. The refractive indexes and densities of the ices used for our calibrations are provided in Table~\ref{tab:calibration}.

\begin{table}[h]
\caption{Refractive indexes and densities used for calibration.}
\begin{center}
\begin{tabular}{c c c c}
\hline
        Molecule & n & $\rho$ (g cm$^{-3}$) & Ref. \\
        \hline
        CH$_{3}$OH & 1.257 & 0.636 & 1 \\
        H$_{2}$O & 1.234 & 0.719 & 2 \\
        CO & 1.297 & 0.870 & 3 \\
    \hline
     %   \noalign{\smallskip}
     \label{tab:calibration}
\end{tabular}

\begin{tablenotes}
\item 1. \citealt{luna2018densities}
\item 2. \citealt{yarnall2022new}
\item 3. \citealt{luna2022density}
\end{tablenotes}

\end{center}
\end{table}

Because the physical properties and masses of H$_{2}$O and HDO are very similar, their flow rates through the leak valves, pumping speeds, and deposition rates are also expected to be very similar, minimizing the expected error of the HDO:H$_{2}$O ice mixing ratio when pre-mixing the components prior to deposition. Therefore, the HDO:H$_{2}$O ices were created via the traditional pre-mixing method, following the procedure described in \cite{galvez2011hdo} to generate HDO via mixing H$_{2}$O and D$_{2}$O.

During the warm-ups, the ices were heated at a rate of 25 K hr$^{-1}$, and spectra ranging from 4000-500 cm$^{-1}$ (2.5-20 $\mu$m) were continuously collected and averaged every 128 scans, meaning a new spectrum was obtained every 3.5 min. This results in an uncertainty of approximately $\pm$1 K in the reported temperature of each amorphous spectrum. During the re-cooling of the annealed crystalline HDO:H$_{2}$O ice mixtures, a cooling rate of -2 K min$^{-1}$ was used so that the experiment could be performed within a single workday, resulting in a higher temperature uncertainty of approximately $\pm$4 K for the annealed crystalline HDO spectra. All spectra were baseline-corrected via cubic spline functions prior to their analysis.

\subsection{HDO O-D stretch characterization}
\begin{figure}
\centering
\includegraphics[width=\linewidth]{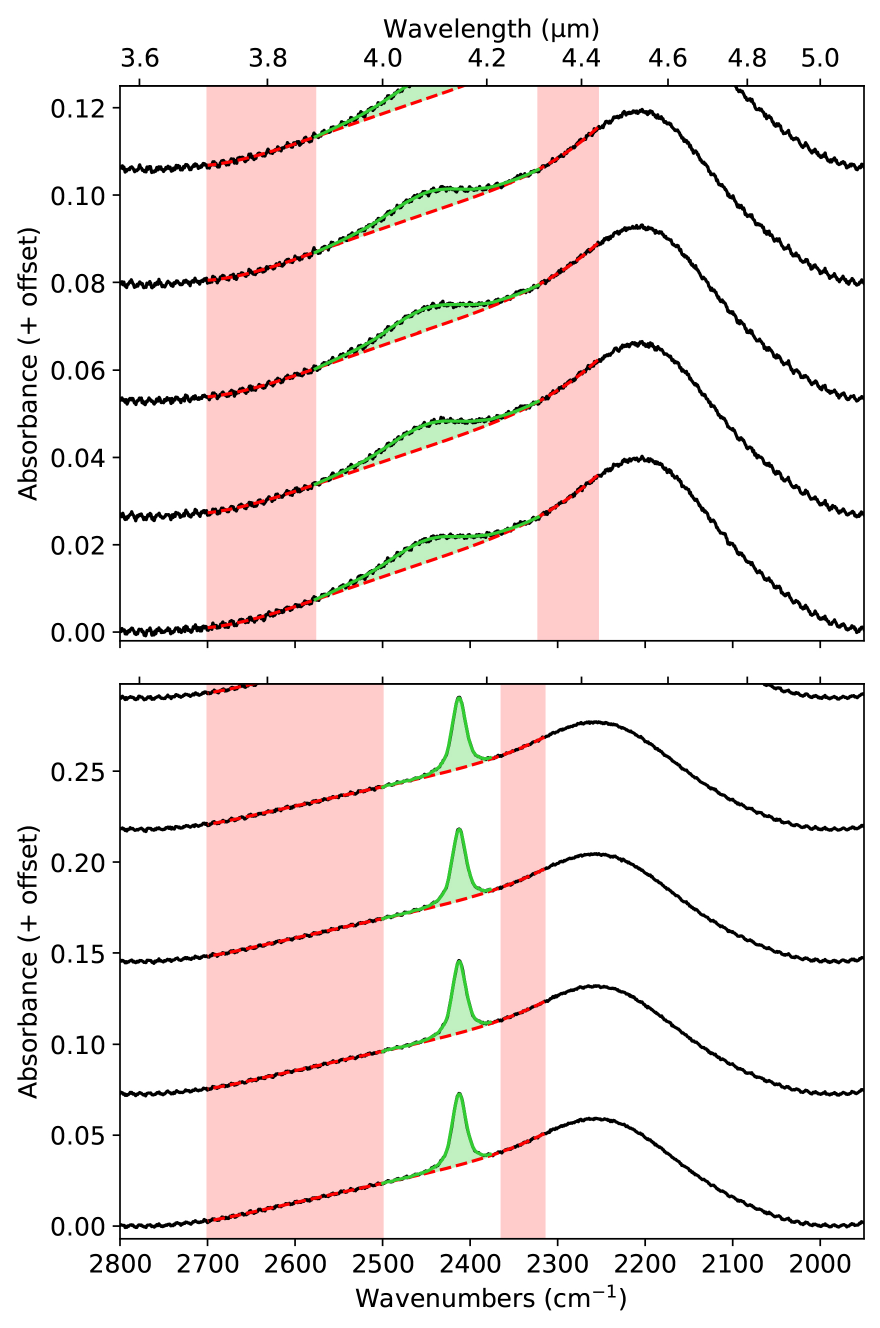}
\caption{Example extractions of the profile of the 4.1 $\mu$m HDO O-D stretching mode from the H$_{2}$O combination mode in 0.4\% HDO:H$_{2}$O laboratory spectra (top: amorphous; bottom: crystalline). The laboratory data (already baseline corrected) is plotted in black, the local continuum is plotted in dashed red, and the smoothed and extracted profile is plotted in green. The red shading indicates the wavelengths used to define the local continuum, and the green shading indicates the integrated area used to derive apparent band strengths.}
\label{fig:hdo_extraction}
\end{figure}

The peak position, FWHM, and integrated absorbance of the HDO O-D stretching mode was extracted from the thick HDO:H$_{2}$O ice spectra (see Table~\ref{tab:lab_spectra}) using a cubic spline to isolate the feature from the broad 4.5 $\mu$m H$_{2}$O combination mode followed by a Savitzky-Golay filter to smooth out experimental noise (see Figure~\ref{fig:hdo_extraction}).

The apparent band strengths of the HDO O-D stretch at different temperatures, $A'_{i}$, were derived by assuming the apparent band strength of the crystalline band at 150 K, $A'_{150K}$, is equal to 6.4$\times$10$^{-17}$ cm molec$^{-1}$ \citep{dartois2003revisiting,galvez2011hdo} and using the approximation:

\begin{equation}
    \indent A'_{i} = \frac{\int \rm abs_{i}(\nu) \ d\nu}{\int \rm abs_{150K}(\nu) \ d\nu} \times A'_{150K},
\label{eq:relbs}
\end{equation}

\noindent where $\int \rm abs_{150K}(\nu) d\nu$ is the integrated absorbance of the crystalline HDO peak at 150 K, and $\int \rm abs_{i}(\nu) d\nu$ is the integrated absorbance of the HDO peak at the phase and temperature of interest.

\begin{figure}[h!]
\centering
\includegraphics[width=\linewidth]{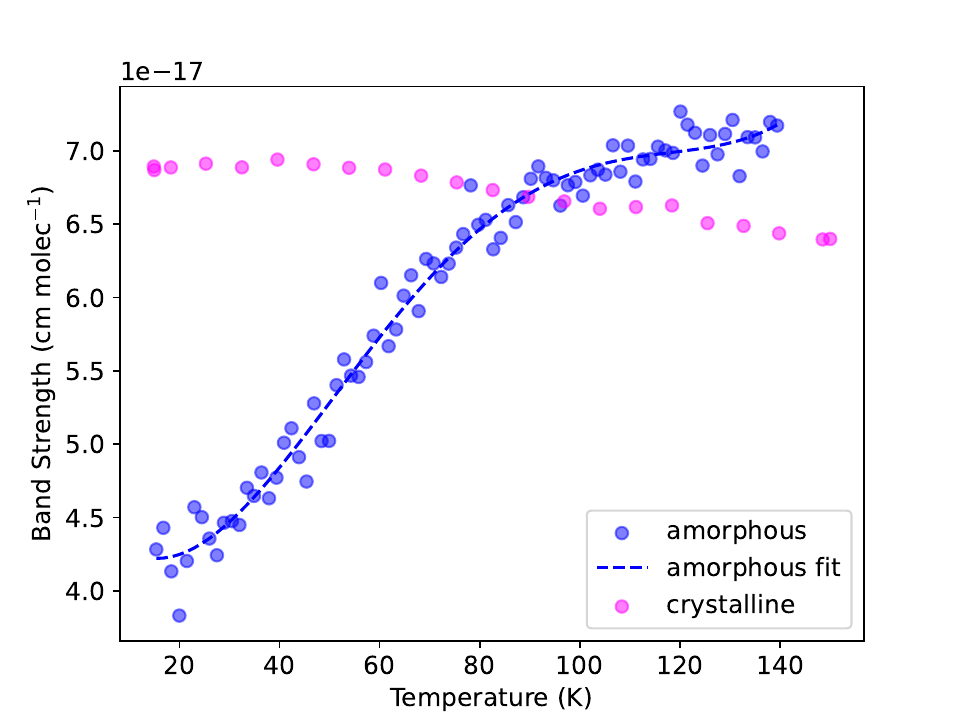}
\caption{The temperature-dependent variation of the apparent band strengths of amorphous and crystalline HDO ice. The dashed blue line is a fourth-order polynomial fit to the amorphous band strength data used to extract the final reported apparent band strengths reported in Table~\ref{tab:hdo_peak_char}.}
\label{fig:hdo_band_strength}
\end{figure}

\begin{figure}[h!]
\centering
\includegraphics[width=\linewidth]{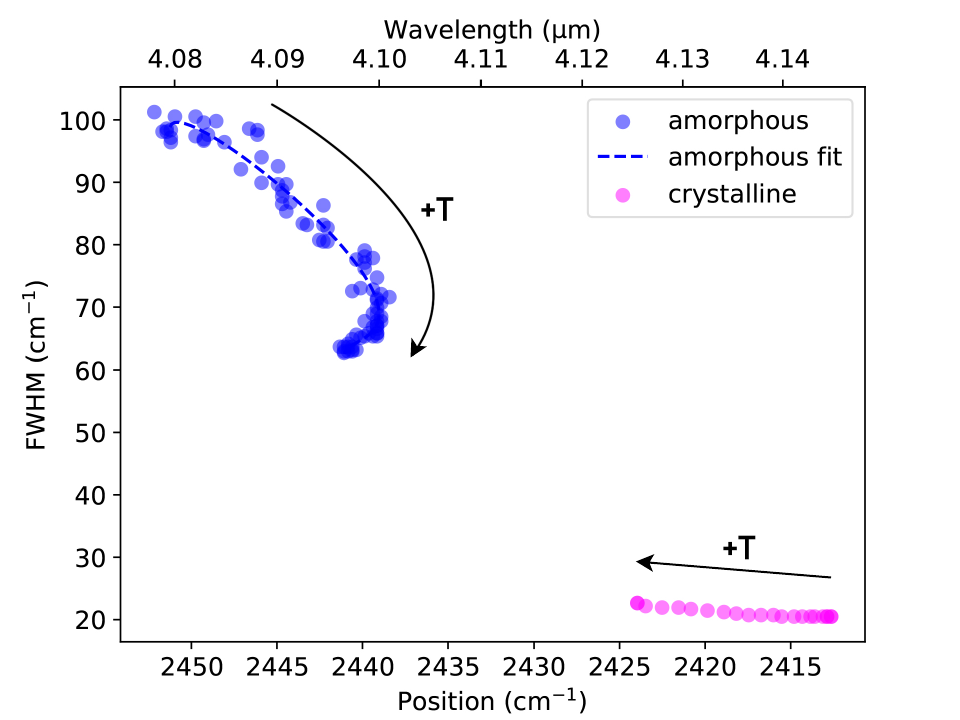}
\caption{The range of peak positions and FWHMs of amorphous and crystalline HDO ice in the heating and annealing experiments. The arrows indicate the direction in which the data changes with increasing temperature.  The dashed blue line is a fourth-order polynomial fit to the amorphous peak position and FWHM data used to extract the final reported peak positions and FWHMs reported in Table~\ref{tab:hdo_peak_char}.}
\label{fig:hdo_pos_fwhm}
\end{figure}

The resulting peak positions, FWHMs, and apparent band strengths are plotted in Figures~\ref{fig:hdo_band_strength} and~\ref{fig:hdo_pos_fwhm} and reported in Table~\ref{tab:hdo_peak_char}. It is evident that the peak extractions are much less reliable for the amorphous ice, where the O-D stretch is very weak and broad (see Figure~\ref{fig:hdo_extraction}), resulting in a greater experimental scatter between the amorphous data points in both plots. We smoothed this scatter with fourth-order polynomial fits to the individual amorphous data points, which were used to calculate the final peak positions, FWHMs, and apparent band strengths reported for the amorphous ice. The crystalline data was not treated in this way because the profile extraction of the crystalline feature from the laboratory spectra was much more straightforward, resulting in much smaller experimental scatter.

Despite the problematic profile of the amorphous HDO ice band, the apparent band strength we derived for amorphous HDO ice at 15 K, 4.2$\times$10$^{-17}$ cm molec$^{-1}$, is in excellent agreement (within $\sim$2\%) with the band strengths reported by \citet{dartois2003revisiting} and \citet{galvez2011hdo} for amorphous HDO ice at similar temperatures.

\begin{table*}[h]
\caption{Peak positions, FWHMs, and band strengths of the HDO O-D stretching mode at $\sim$4.1 $\mu$m at temperatures from 15-150 K. We estimate uncertainties of 2 cm$^{-1}$ for the peak positions/FWHMs and 20\% for the band strengths of the amorphous HDO ice, and 0.5 cm$^{-1}$ for the peak positions/FWHMs and 10\% for the band strengths of the crystalline HDO ice.}
\begin{center}
\begin{tabular}{|c|c|cc|cc|c|}
\hline
    \multirow{2}{*}{Ice phase} & \multirow{2}{*}{T (K)} & \multicolumn{2}{c|}{Peak} & \multicolumn{2}{c|}{FWHM} & $A'_{i}$ \\
    &  & (cm$^{-1}$) & ($\mu$m) & (cm$^{-1}$) & ($\mu$m) & (10$^{-17}$ cm molec$^{-1}$) \\
    \hline 
    \multirow{13}{*}{Amorphous} & 15 & 2451.3 & 4.080 & 98.7 & 0.164 & 4.2 \\
     & 25 & 2450.5 & 4.081 & 99.5 & 0.166 & 4.3 \\
     & 35 & 2448.8 & 4.084 & 97.3 & 0.162 & 4.6 \\
     & 45 & 2446.5 & 4.087 & 93.2 & 0.156 & 5.1 \\
     & 55 & 2444.3 & 4.091 & 88.1 & 0.147 & 5.5 \\
     & 65 & 2442.2 & 4.095 & 82.7 & 0.139 & 5.9 \\
     & 75 & 2440.6 & 4.097 & 77.6 & 0.130 & 6.3 \\
     & 85 & 2439.5 & 4.099 & 73.2 & 0.123 & 6.6 \\
     & 95 & 2439.1 & 4.100 & 69.7 & 0.117 & 6.8 \\
     & 105 & 2439.2 & 4.100 & 67.2 & 0.113 & 6.9 \\
     & 115 & 2439.7 & 4.099 & 65.5 & 0.110 & 7.0 \\
     & 125 & 2440.4 & 4.098 & 64.3 & 0.108 & 7.0 \\
     & 135 & 2440.9 & 4.097 & 63.1 & 0.106 & 7.1 \\
    \hline
    \multirow{8}{*}{Crystalline} & 15 & 2412.6 & 4.145 & 20.5 & 0.035 & 6.9 \\
    & 25 & 2412.9 & 4.144 & 20.5 & 0.035 & 6.9 \\
    & 47 & 2413.8 & 4.143 & 20.5 & 0.035 & 6.9 \\
    & 68 & 2415.5 & 4.140 & 20.5 & 0.035 & 6.8 \\
    & 90 & 2417.5 & 4.137 & 20.7 & 0.035 & 6.7 \\
    & 111 & 2419.9 & 4.132 & 21.5 & 0.037 & 6.6 \\
    & 133 & 2422.5 & 4.128 & 21.9 & 0.037 & 6.5 \\
    & 150 & 2424.0 & 4.125 & 22.7 & 0.039 & 6.4 \\
    \hline
     %   \noalign{\smallskip}
\end{tabular}
\end{center}
\label{tab:hdo_peak_char}
\end{table*}

\subsection{CH$_{3}$OH C-H symmetric stretch characterization}
\begin{figure}[H]
\centering
\includegraphics[width=\linewidth]{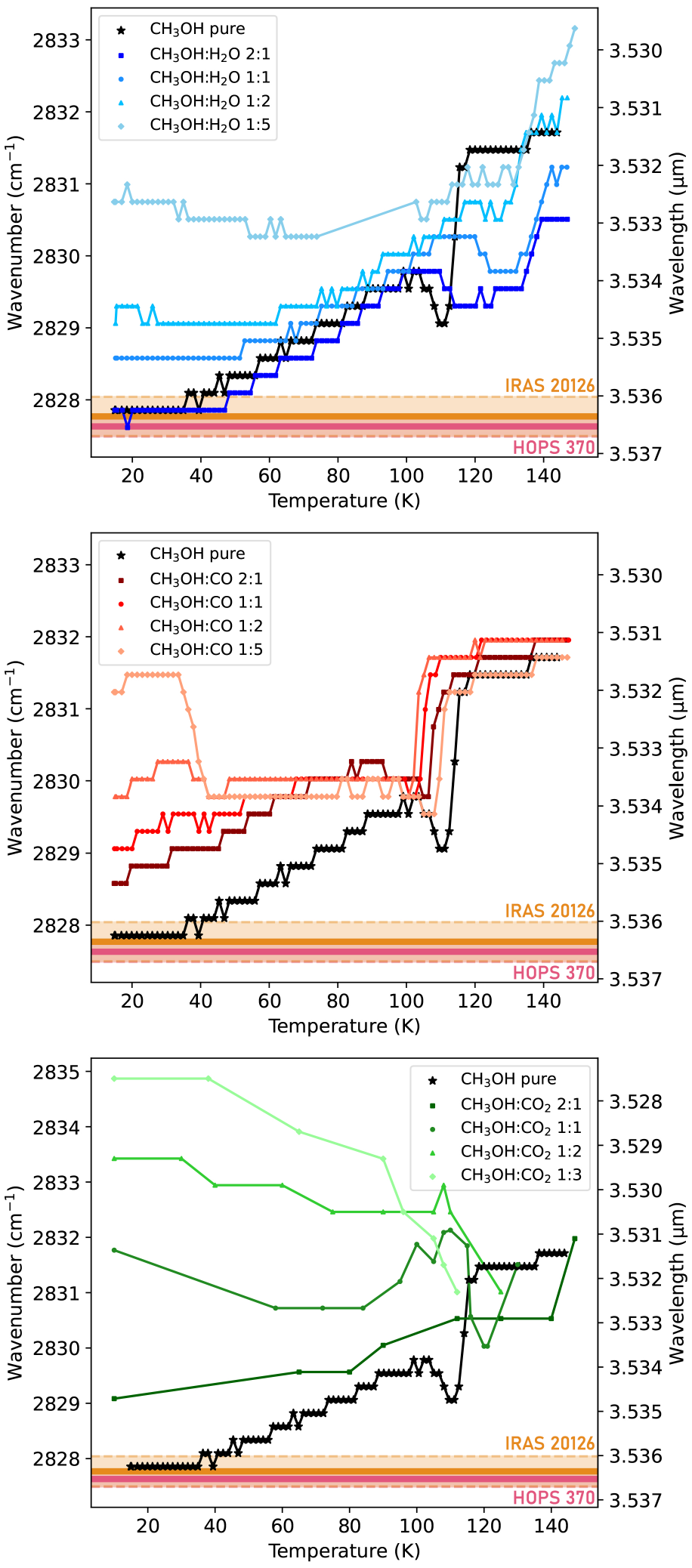}
\caption{Peak positions of the 3.53 $\mu$m CH$_{3}$OH C-H stretching mode in laboratory CH$_{3}$OH ices from 15-145 K (points), and peak positions of the 3.53 $\mu$m band observed toward HOPS 370 and IRAS 20126 (thick horizontal lines). The observed peak positions were extracted from the observed spectra via Gaussian fits to the local-continuum subtracted 3.53 $\mu$m peaks (error margins indicated with shaded areas with dashed borders). The CH$_{3}$OH:CO$_{2}$ values were extracted from data from \cite{ehrenfreund1999laboratory} available on LIDA.}
\label{fig:ch3oh_pos}
\end{figure}

The peak position of the CH$_{3}$OH C-H symmetric stretching mode was extracted from the CH$_{3}$OH-containing ice spectra using a fourth-order polynomial to isolate the feature from the entire 3.4 $\mu$m absorption complex followed by a Savitzky-Golay filter to smooth out experimental noise. These values are given in Tables~\ref{tab:ch3oh_peak_char_h2o} and~\ref{tab:ch3oh_peak_char_co} and are plotted as a function of temperature in Figure~\ref{fig:ch3oh_pos}, which also shows the observed peak positions of the C-H symmetric stretching mode in the analyzed spectra. Only peak positions are provided here because a full characterization of this peak (i.e., including FWHMs and relative apparent band strengths) is complicated by the fact that this feature is blended with the CH$_{3}$OH overtones and C-H asymmetric stretching modes.

\begin{table*}[h!]
\caption{Peak positions of the CH$_{3}$OH C-H symmetric stretching mode at $\sim$3.53 $\mu$m at temperatures from 15-145 K in pure CH$_{3}$OH ice and CH$_{3}$OH ice mixed with H$_{2}$O. We estimate uncertainties of 0.5 cm$^{-1}$ for these peak positions.}
\begin{center}
\begin{tabular}{|c|cc|cc|cc|cc|cc|}
\hline
    \multirow{2}{*}{T (K)} & \multicolumn{2}{c|}{Pure CH$_{3}$OH} & \multicolumn{2}{c|}{CH$_{3}$OH:H$_{2}$O 2:1} & \multicolumn{2}{c|}{CH$_{3}$OH:H$_{2}$O 1:1} & \multicolumn{2}{c|}{CH$_{3}$OH:H$_{2}$O 1:2} & \multicolumn{2}{c|}{CH$_{3}$OH:H$_{2}$O 1:5} \\
    & (cm$^{-1}$) & ($\mu$m) & (cm$^{-1}$) & ($\mu$m) & (cm$^{-1}$) & ($\mu$m) & (cm$^{-1}$) & ($\mu$m) & (cm$^{-1}$) & ($\mu$m) \\
    \hline
    15 & 2827.9 & 3.5362 & 2827.9 & 3.5362 & 2828.6 & 3.5353 & 2829.1 & 3.5347 & 2830.7 & 3.5326 \\
    25 & 2827.9 & 3.5362 & 2827.9 & 3.5362 & 2828.6 & 3.5353 & 2829.1 & 3.5347 & 2830.7 & 3.5326 \\
    35 & 2827.9 & 3.5362 & 2827.9 & 3.5362 & 2828.6 & 3.5353 & 2829.1 & 3.5347 & 2830.7 & 3.5326 \\
    45 & 2828.3 & 3.5366 & 2827.9 & 3.5362 & 2828.6 & 3.5353 & 2829.1 & 3.5347 & 2830.5 & 3.5329 \\
    55 & 2828.3 & 3.5366 & 2828.3 & 3.5356 & 2828.8 & 3.5350 & 2829.1 & 3.5347 & 2830.3 & 3.5332 \\
    65 & 2828.6 & 3.5353 & 2828.6 & 3.5353 & 2828.8 & 3.5350 & 2829.3 & 3.5344 & 2830.3 & 3.5332 \\
    75 & 2829.1 & 3.5347 & 2828.8 & 3.5350 & 2829.3 & 3.5344 & 2829.5 & 3.5341 & 2830.3 & 3.5332 \\
    85 & 2829.3 & 3.5344 & 2829.1 & 3.5347 & 2829.5 & 3.5341 & 2829.5 & 3.5341 & - & - \\
    95 & 2829.5 & 3.5341 & 2829.5 & 3.5341 & 2829.8 & 3.5338 & 2830.0 & 3.5335 & - & -\\
    105 & 2829.5 & 3.5341 & 2829.8 & 3.5338 & 2830.0 & 3.5335 & 2830.3 & 3.5332 & 2830.5 & 3.5329 \\
    115 & 2831.2 & 3.5320 & 2829.3 & 3.5344 & 2830.3 & 3.5332 & 2830.5 & 3.5329 & 2831.0 & 3.5323 \\
    125 & 2831.5 & 3.5317 & 2829.3 & 3.5344 & 2829.8 & 3.5338 & 2830.7 & 3.5326 & 2831.0 & 3.5323 \\
    135 & 2831.5 & 3.5317 & 2829.8 & 3.5338 & 2830.0 & 3.5335 & 2831.7 & 3.5314 & 2831.7 & 3.5314 \\
    145 & 2831.7 & 3.5314 & 2830.5 & 3.5329 & 2831.2 & 3.5320 & 2832.2 & 3.5308 & 2832.7 & 3.5302 \\
    \hline
     %   \noalign{\smallskip}
\end{tabular}
\end{center}
\label{tab:ch3oh_peak_char_h2o}
\end{table*}

\clearpage

\begin{table*}[h!]
\caption{Peak positions of the CH$_{3}$OH C-H symmetric stretching mode at $\sim$3.53 $\mu$m at temperatures from 15-145 K in CH$_{3}$OH ice mixed with CO. We estimate uncertainties of 0.5 cm$^{-1}$ for these peak positions.}
\begin{center}
\begin{tabular}{|c|cc|cc|cc|cc|}
\hline
    \multirow{2}{*}{T (K)} & \multicolumn{2}{c|}{CH$_{3}$OH:CO 2:1} & \multicolumn{2}{c|}{CH$_{3}$OH:CO 1:1} & \multicolumn{2}{c|}{CH$_{3}$OH:CO 1:2} & \multicolumn{2}{c|}{CH$_{3}$OH:CO 1:5} \\
    & (cm$^{-1}$) & ($\mu$m) & (cm$^{-1}$) & ($\mu$m) & (cm$^{-1}$) & ($\mu$m) & (cm$^{-1}$) & ($\mu$m) \\
    \hline
    15 & 2828.6 & 3.5353 & 2829.1 & 3.5347 & 2829.8 & 3.5338 & 2831.2 & 3.5320 \\
    25 & 2828.8 & 3.5350 & 2829.3 & 3.5344 & 2830.0 & 3.5335 & 2831.5 & 3.5317 \\
    35 & 2829.1 & 3.5347 & 2829.5 & 3.5341 & 2830.3 & 3.5332 & 2831.2 & 3.5320 \\
    45 & 2829.1 & 3.5347 & 2829.5 & 3.5341 & 2829.8 & 3.5338 & 2829.8 & 3.5338 \\
    55 & 2829.5 & 3.5341 & 2829.8 & 3.5338 & 2830.0 & 3.5335 & 2829.8 & 3.5338 \\
    65 & 2829.8 & 3.5338 & 2829.8 & 3.5338 & 2830.0 & 3.5335 & 2829.8 & 3.5338 \\
    75 & 2830.0 & 3.5335 & 2830.0 & 3.5335 & 2830.0 & 3.5335 & 2829.8 & 3.5338 \\
    85 & 2830.0 & 3.5335 & 2830.0 & 3.5335 & 2830.0 & 3.5335 & 2829.8 & 3.5338 \\
    95 & 2830.0 & 3.5335 & 2830.0 & 3.5335 & 2830.0 & 3.5335 & 2829.8 & 3.5338 \\
    105 & 2829.8 & 3.5338 & 2831.0 & 3.5323 & 2831.5 & 3.5317 & 2829.5 & 3.5341 \\
    115 & 2831.5 & 3.5317 & 2831.7 & 3.5314 & 2831.7 & 3.5314 & 2831.2 & 3.5320 \\
    125 & 2831.7 & 3.5314 & 2832.0 & 3.5311 & 2832.0 & 3.5311 & 2831.5 & 3.5317 \\
    135 & 2831.7 & 3.5314 & 2832.0 & 3.5311 & 2832.0 & 3.5311 & 2831.5 & 3.5317 \\
    145 & 2832.0 & 3.5311 & 2832.0 & 3.5311 & 2832.0 & 3.5311 & 2831.7 & 3.5314 \\
    \hline
     %   \noalign{\smallskip}
\end{tabular}
\end{center}
\label{tab:ch3oh_peak_char_co}
\end{table*}

\clearpage

\section{$\chi^{2}$ plots}
\label{app:chi2}

\begin{figure}[H]
\centering
\includegraphics[width=\linewidth]{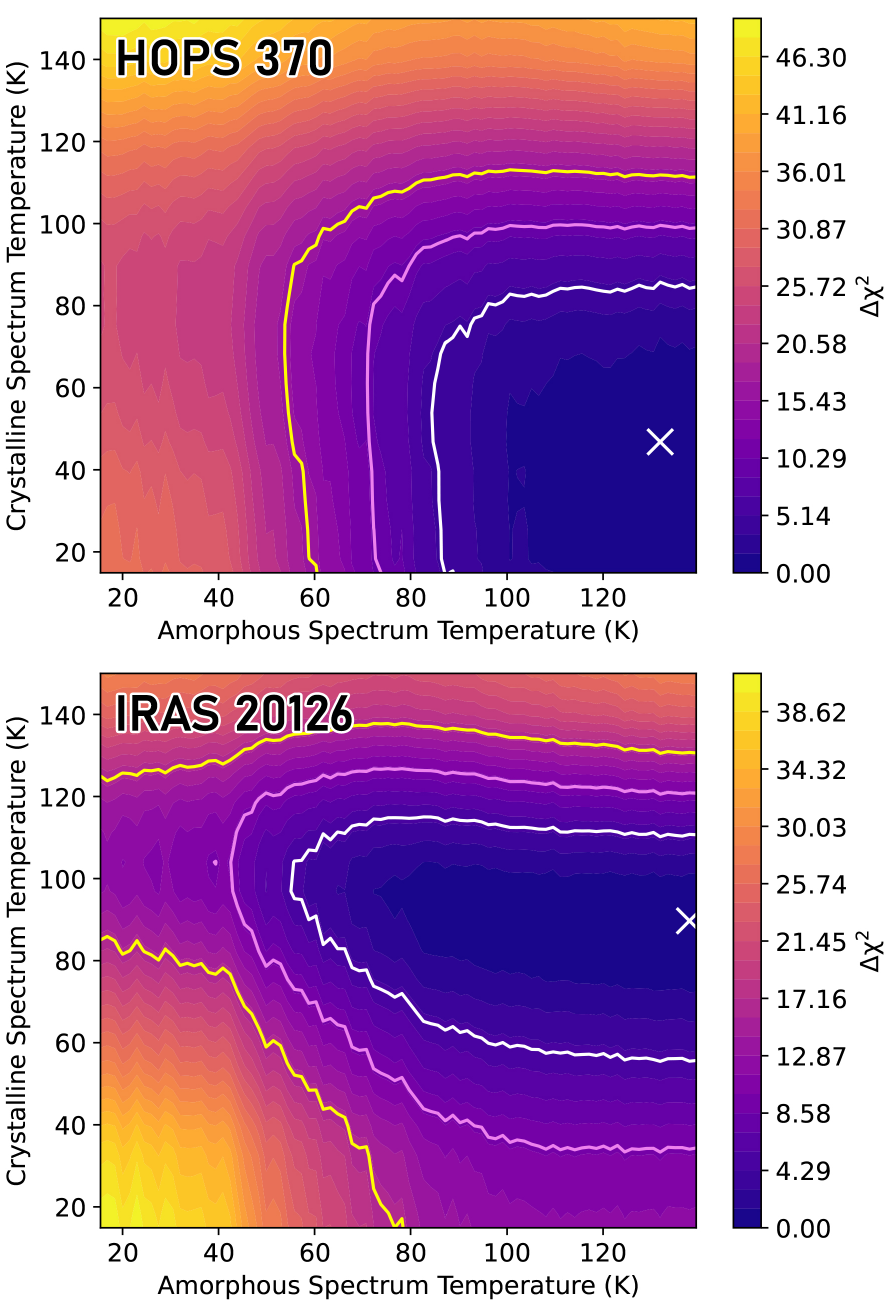}
\caption{Contour maps of the $\Delta$$\chi^{2}$ values of the least-squares fits with every possible combination of one amorphous and one crystalline HDO component. The white X marks $\chi^{2}$$_{min}$, and the white, purple, and yellow contour lines show the $\Delta$$\chi^{2}$ values corresponding to 1, 2, and 3$\sigma$, respectively.}
\label{fig:chi2_plots}
\end{figure}

Least-squares fits using every possible combination of one amorphous and one crystalline HDO lab spectra were performed, and the $\chi^{2}$ values were calculated for all of these possible fits using the following equation:

\begin{equation}
    \indent \chi^{2} = \frac{\sum(\tau_{obs} - \tau_{fit})^{2}}{\sigma^{2}},
\end{equation}

\noindent where $\tau_{obs}$ is the observed optical depth and $\tau_{fit}$ is the optical depth of the fit. The $\sigma$ value is calculated by propagating the uncertainties in the optical depths that result from the RMS errors and the local continuum placement uncertainties (Figures~\ref{fig:hdo_contsub_var} and~\ref{fig:hdo_contsub_dev}). The 1, 2, and 3$\sigma$ bounds plotted in Figure~\ref{fig:chi2_plots} correspond to the critical $\Delta$$\chi^{2}$ ($\chi^{2}$-$\chi^{2}$$_{min}$) values at the 68.3, 95.5, and 99.7\% confidence intervals, respectively, for fits with four free parameters (in this case, the temperatures and scaling factors of the amorphous and crystalline HDO ice spectra).

\clearpage

\section{Estimated continuum errors}
\label{app:cont_err}

\subsection{CH$_{3}$OH ice}
\label{app:cont_err_ch3oh}
The 20\% band strength uncertainty \citep{luna2018densities} is propagated to determine the uncertainty in the calculated CH$_{3}$OH ice column densities. The implications of these fits and column densities on the thermal and chemical structures of the ice envelopes as well as the physicochemical histories of the sources are discussed in Section~\ref{txt:ch3oh_ice_formation}.

\subsection{HDO ice}
\label{app:cont_err_hdo}
The reported errors account for two sources of uncertainty: 1) the uncertainty of the HDO O-D stretching mode band strength, and 2) the uncertainty of the local continuum choice. The uncertainty in the band strength accounts for both the aforementioned variation of the band strength with temperature as well as the reported experimental uncertainties of the amorphous and crystalline HDO band strengths from \cite{galvez2011hdo} ($\sim$10\% and 5\%, respectively), which were used to calculate the band strengths of HDO at higher temperatures presented in Table~\ref{tab:hdo_peak_char}.

\begin{figure}
\centering
\includegraphics[width=\linewidth]{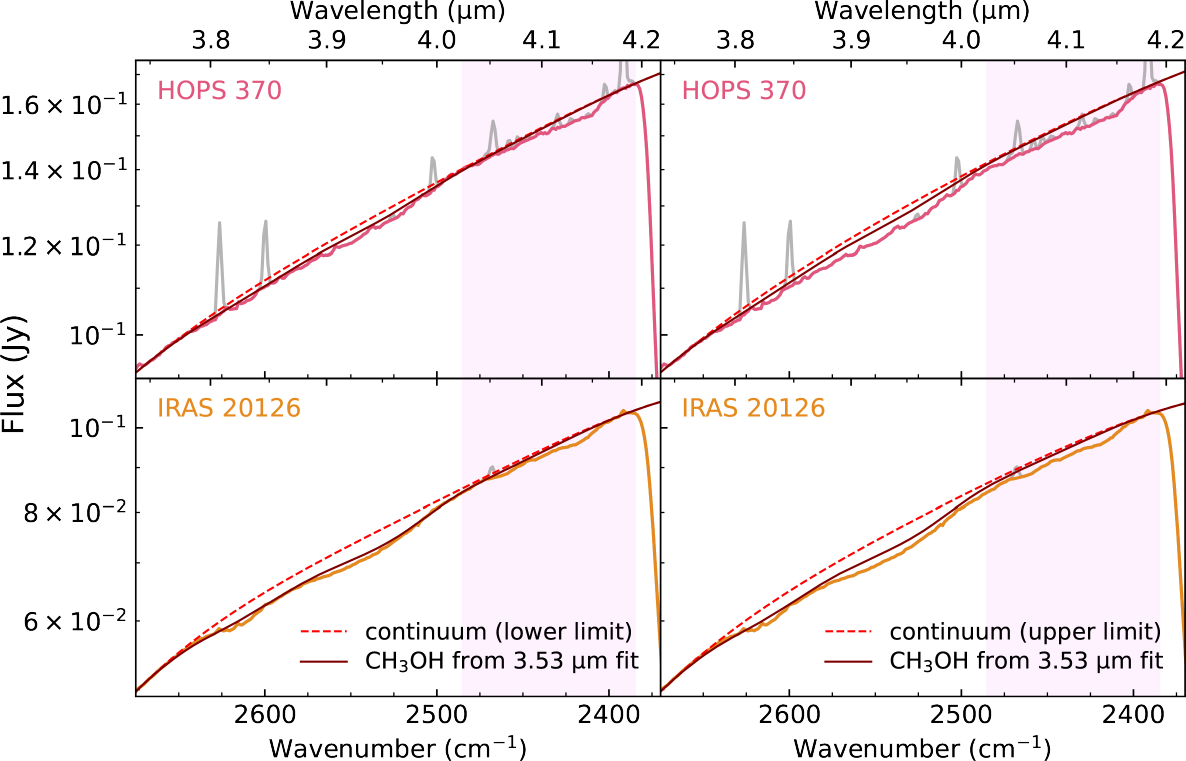}
\caption{The alternative local continua used to calculate the uncertainty in HDO column density from local continuum choice. Left: the lower limit continua (i.e., the continua that result in the lowest possible optical depth in the region while remaining reasonable). Right: the upper limit continua (i.e., the continua that result in the highest possible optical depth in the region while remaining reasonable).}
\label{fig:hdo_contsub_var}
\end{figure}

\begin{figure}
\centering
\includegraphics[width=\linewidth]{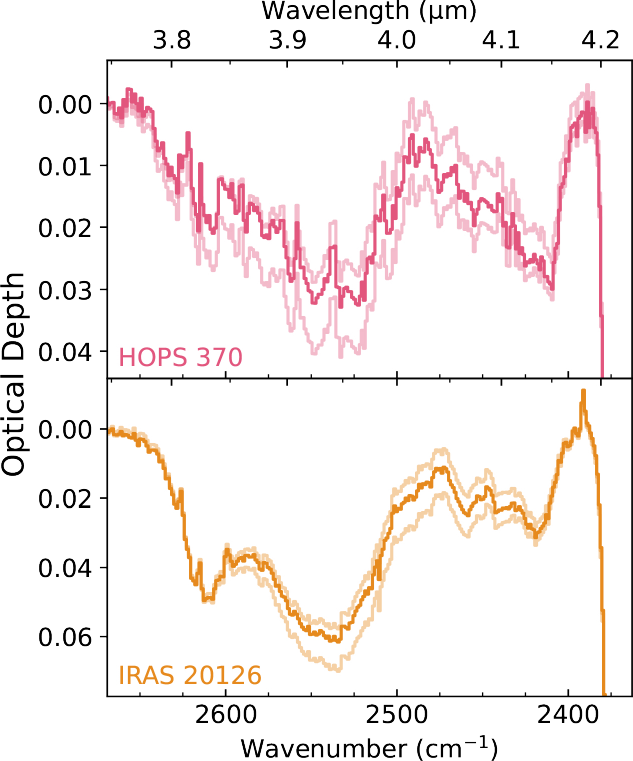}
\caption{The change the optical depth of the features at 4 $\mu$m caused by varying the local continua as shown in Figure~\ref{fig:hdo_contsub_var}. The spectra used to calculate the column densities in Table~\ref{tab:column_densities_hdo} are plotted in the dark shades, and the spectra resulting from the lower and upper continuum limits used to estimate the effect of continuum uncertainty are plotted in the light shades.}
\label{fig:hdo_contsub_dev}
\end{figure}

The uncertainty of the local continuum is somewhat subjective, as it depends on what one considers a "reasonable" local continuum. We attempted to quantify this uncertainty by identifying the limits of what we consider a reasonable local continuum for each line of sight and then quantifying the variation in the calculated HDO column density that resulted from these continuum shifts. These lower and upper limit continua are shown in Figure~\ref{fig:hdo_contsub_var}, and the resulting lower and upper limits on the optical depths of the 4 $\mu$m region are shown in Figure~\ref{fig:hdo_contsub_dev}.

The column density uncertainties were determined by adding in quadrature the band strength uncertainties and the uncertainties caused by the continuum choice, which were estimated by noting the variation in HDO ice column density obtained from repeating the fitting procedure on the spectra that were extracted using the lower and upper limit global continua.

\subsection{H$_{2}$O ice}
\label{app:cont_err_h2o}
\begin{figure}[H]
\centering
\includegraphics[width=\linewidth]{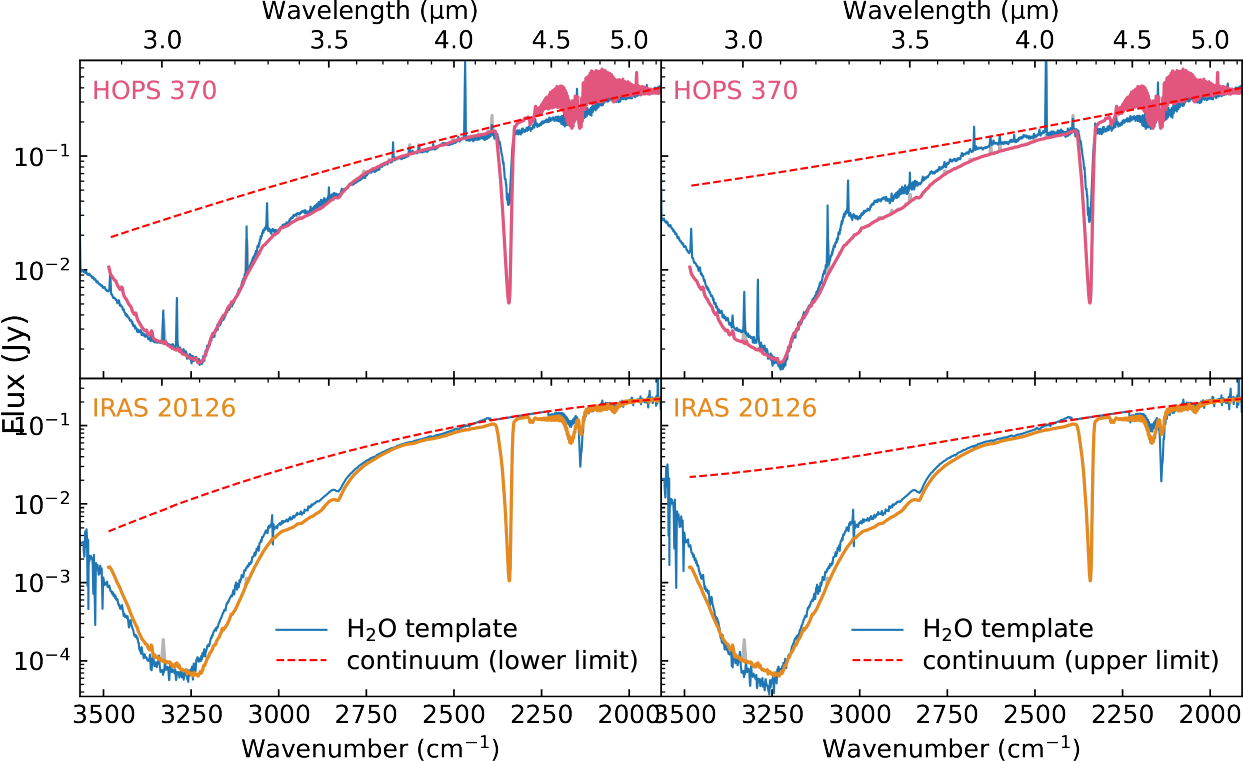}
\caption{The alternative global continua used to calculate the uncertainty in H$_{2}$O column density from local continuum choice. Left: the lower limit continua (i.e., the continua that result in the lowest possible optical depth in the region while remaining reasonable). Right: the upper limit continua (i.e., the continua that result in the highest possible optical depth in the region while remaining reasonable).}
\label{fig:h2o_contsub_var}
\end{figure}

\begin{figure}[H]
\centering
\includegraphics[width=\linewidth]{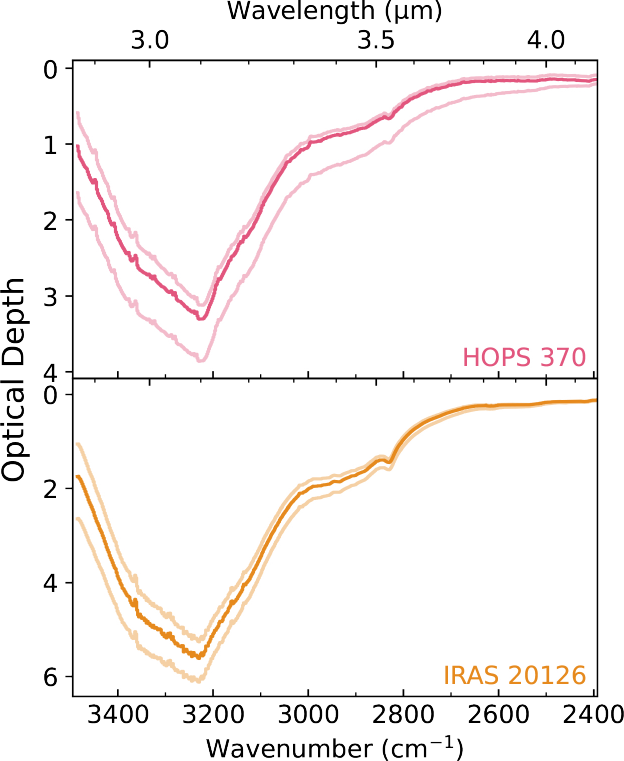}
\caption{The change in the optical depth of the 3 $\mu$m feature caused by varying the global continua as shown in Figure~\ref{fig:h2o_contsub_var}. The spectra used to calculate the column densities in Table~\ref{tab:column_densities_h2o} are plotted in the dark shades, and the spectra resulting from the lower and upper continuum limits used to estimate the effect of continuum uncertainty are plotted in the light shades.}
\label{fig:h2o_contsub_dev}
\end{figure}

The uncertainty in the H$_{2}$O column density due to the uncertainty of the continuum choice was quantified via a procedure similar to that used for the 4 $\mu$m region by exploring the range of artificial short-wavelength continuum points which resulted in continuum shapes and H$_{2}$O profiles that appeared reasonable. The lower and upper limit continua are shown in Figure~\ref{fig:h2o_contsub_var}, and the resulting variations in the optical depths and profiles of the 3 $\mu$m feature are shown in Figure~\ref{fig:h2o_contsub_dev}.

The column density uncertainties were determined by adding in quadrature the uncertainties caused by the variation in band strengths with temperature, a 30\% uncertainty accounting for the underfit red wing caused by neglecting ammonia hydrates and grain shape effects (as estimated by \citealt{boogert2008c2d}), and the uncertainties caused by the continuum choice, which were estimated by noting the variation in H$_{2}$O ice column density obtained from repeating the fitting procedure on the spectra that were extracted using the lower and upper limit global continua.

\clearpage

\newpage

\section{Literature HDO/H$_{2}$O values}
\label{app:hdo_h2o_lit}

\begin{table}[h]
\caption{Overview of recently published HDO/H$_{2}$O values in ices and the gas phase (adapted from \citealt{jensen2019alma} and \citealt{andreu2023high}).}
\begin{center}
\begin{tabular}{l c c c}
\hline
        Source & Phase & HDO/H$_{2}$O & Ref. \\
        & & 10$^{-3}$ & \\
        \hline
        \multicolumn{4}{c}{Clustered Class 0 LYSOs} \\
        \hline
        NGC 1333 IRAS 4A-NW & gas & 0.54$\pm$0.15 & 1, 2 \\
        NGC 1333 IRAS 2A & gas & 0.74$\pm$0.21 & 1 \\
        NGC 1333 IRAS 4B & gas & 0.59$\pm$0.26 & 1 \\
        IRAS 16293-2422 & gas & 0.92$\pm$0.26 & 1 \\
        NGC 1333 SVS 13 & ice & $\leq$17 & 3 \\
        \hline
        \multicolumn{4}{c}{Clustered Class I LYSOs} \\
        \hline
        NGC 1333 SVS 12 & ice & $\leq$5 & 3 \\
        \hline
        \multicolumn{4}{c}{Isolated Class 0 LYSOs} \\
        \hline
        BHR 71-IRS1 & gas & 1.8$\pm$0.4 & 2 \\
        B335 & gas & 6.3$\pm$1.5 & 4 \\
        L483 & gas & 4.0$\pm$0.5 & 4 \\
        \hline
        \multicolumn{4}{c}{Isolated Class I LYSOs} \\
        \hline
        V883 Ori & gas & 2.3$\pm$0.6 & 5 \\
        L1551 IRS5 & gas & 2.1$\pm$0.8 & 6 \\
        L1489 IRS & ice & $\leq$8 & 3 \\
        TMR1 & ice & $\leq$11 & 3 \\
        \hline
        \multicolumn{4}{c}{IMYSOs} \\
        \hline
        IRAS 05390-0728 & ice & $\leq$10 & 7 \\
        IRAS 08448-4343 & ice & $\leq$10 & 7 \\
        HOPS 370 & ice & 4.6$\pm$2.2 & 8 \\
        \hline
        \multicolumn{4}{c}{MYSOs} \\
        \hline
        W3 IRS5 & gas & 1.3 & 9 \\
        W33A & gas & 3.0 & 9 \\
        AFGL 2591 & gas & 3.3 & 9 \\
        NGC 7538 IRS1 & gas & 3.8 & 9 \\
        Orion KL Hot Core & gas & 3.0$^{+3.1}_{-1.7}$ & 10 \\
        NGC 6334 I & gas & 0.21$\pm$0.1 & 11 \\
        G34.26+0.15 & gas & 0.35-0.75 & 12 \\
        NGC 7538 IRS9 & ice & $\leq$8.1-11.4 & 7 \\
        GL 2136 & ice & $\leq$4 & 7 \\
        IRAS 20126 & ice & 2.6$\pm$1.4 & 8 \\
    \hline
     %   \noalign{\smallskip}
     \label{tab:hdo_h2o_lit}
\end{tabular}

\begin{tablenotes}
    \item 1. \citealt{persson2014deuterium}
    \item 2. \citealt{jensen2019alma}
    \item 3. \citealt{parise2003search}
    \item 4. \citealt{jensen2021alma}
    \item 5. \citealt{tobin2023deuterium}
    \item 6. \citealt{andreu2023high}
    \item 7. \citealt{dartois2003revisiting}
    \item 8. this work
    \item 9. \citealt{van2006water} (envelope-averaged abundances from radiative transfer models)
    \item 10. \citealt{neill2013abundance}
    \item 11. \citealt{emprechtinger2013abundance}
    \item 12. \citealt{coutens2014water}
\end{tablenotes}

\end{center}
\end{table}

\end{appendix}

\end{document}